\documentclass{article}

\usepackage[preprint]{neurips_2025}

\usepackage[utf8]{inputenc} 
\usepackage[T1]{fontenc}    
\usepackage{hyperref}       
\usepackage{url}            
\usepackage{booktabs}       
\usepackage{amsfonts}       
\usepackage{nicefrac}       
\usepackage{microtype}      
\usepackage{xcolor}         
\usepackage{cancel}
\usepackage{amsmath}
\usepackage{amssymb}
\usepackage{amsthm}
\usepackage{graphicx}
\usepackage{subcaption}
\usepackage{tikz}
\usetikzlibrary{quantikz2}
\usepackage{verbatim}
\usepackage{placeins}
\newtheorem{theorem}{Theorem}

\renewcommand{\vec}[1]{\boldsymbol{#1}}

\title{Spectral Bias in Variational Quantum Machine Learning}

\author{
  Callum Duffy \\
  Department of Physics and Astronomy\\
  Centre for Data Intensive Science and Industry\\
  University College London\\
  Gower Street, London WC1E 6BT, United Kingdom\\
  \texttt{callum.duffy.22@ucl.ac.uk} \\
  \And
  Marcin Jastrzebski \\
  Department of Physics and Astronomy\\
  University College London\\
  Gower Street, London WC1E 6BT, United Kingdom\\
  \texttt{marcin.jastrzebski.21@ucl.ac.uk} \\
}

\begin{document}

\maketitle

\begin{abstract}
In this work, we investigate the phenomenon of spectral bias in quantum machine learning, where, in classical settings, models tend to fit low-frequency components of a target function earlier during training than high-frequency ones, demonstrating a frequency-dependent rate of convergence. We study this effect specifically in parameterised quantum circuits (PQCs). Leveraging the established formulation of PQCs as Fourier series, we prove that spectral bias in this setting can arise from the ``redundancy'' of the Fourier coefficients, which denotes the number of terms in the analytical form of the model contributing to the same frequency component. The choice of data encoding scheme dictates the degree of redundancy for a Fourier coefficient.  We then further demonstrate this empirically with several different encoding schemes. Additionally, we demonstrate that PQCs with greater redundancy exhibit increased robustness to random perturbations in their parameters at the corresponding frequencies. We investigate how design choices affect the ability of PQCs to learn Fourier sums, focusing on parameter initialization scale and entanglement structure, finding large initializations and low-entanglement schemes tend to slow convergence.
\end{abstract}

\section{Introduction}
Parameterised quantum circuits (PQCs) are a class of machine learning models often utilised in quantum machine learning (QML), one of the most promising applications of quantum computing \citet{wiebe2015quantumdeeplearning}. These models resemble classical neural networks in that they consist of trainable quantum gates whose parameters are optimized via hybrid quantum-classical algorithms. Classical data can be encoded into the quantum circuit similarly, through parametrised gates. As with deep neural networks (DNNs), developing a theoretical understanding of PQCs is essential for advancing their practical capabilities and understanding their limitations. In classical settings, it is well known that DNNs exhibit spectral bias, a tendency to learn low frequency functions more readily than high frequency ones \cite{cao2020understandingspectralbiasdeep}. This bias may elucidate why large neural networks have demonstrated low generalization error \citet{cao2020understandingspectralbiasdeep, xu2018understandingtraininggeneralizationdeep, xu2024overviewfrequencyprinciplespectralbias}. However, in certain problems, models capable of capturing high-frequency components in the data have been found to be beneficial. Such instances can be found in domains of image recognition \citet{fang2024inexactfppaell0sparse, Lu_2010} and solutions to PDEs \citet{krishnapriyan2021characterizingpossiblefailuremodes, wang2020understandingmitigatinggradientpathologies}. Strides have been taken to mitigate the effects of spectral bias in deep neural networks \citet{Ziqi_Liu_2020, JAGTAP2020109136, cai2019phaseshiftdeepneural, tancik2020fourierfeaturesletnetworks, fang2024addressingspectralbiasdeep}. Recent work has shown that PQC outputs are expressible as Fourier series, with the data encoding strategy governing the frequency spectrum accessible to the circuit \citet{PhysRevA.103.032430}. In this work, we leverage this framework to study the training dynamics of PQCs in learning target functions with specific frequency content.

\section{Related Work}
Since the relation between PQCs and Fourier series was established in \cite{PhysRevA.103.032430}, it has been a powerful framework to understand their expressivity and properties. Demonstrating that PQCs following a reuploader scheme are universal function approximators \cite{PhysRevA.103.032430, P_rez_Salinas_2020}. Certain studies have aimed to explore how the choice of data encoding affects the spectrum \cite{PhysRevA.107.012422} as well as how the encoding subsequently impacts on the generalization bounds \cite{Caro_2021}. Further work has looked into the learning capabilities of PQCs from a Fourier perspective \cite{heimann2024learningcapabilityparametrizedquantum}. This framework is used thoroughly in the search for quantum advantage in PQCs, shedding light on conditions under which PQCs can be dequantized \cite{Sweke_2025, landman2022classicallyapproximatingvariationalquantum}.

In parallel to this, the classical machine learning community has extensively studied spectral bias. This phenomenon has been demonstrated in fully connected networks \cite{xu2018understandingtraininggeneralizationdeep, cao2020understandingspectralbiasdeep}, convolutional \cite{xu2024overviewfrequencyprinciplespectralbias} and physics-informed neural networks \cite{krishnapriyan2021characterizingpossiblefailuremodes}. Theoretical results have supported these findings, connecting spectral bias to neural tangent kernels, Fourier feature mappings and gradient flow \cite{geifman2022spectralbiasconvolutionalneural, cao2020understandingspectralbiasdeep, tancik2020fourierfeaturesletnetworks, bordelon2021spectrumdependentlearningcurves, basri2019convergencerateneuralnetworks}. Motivated by the limitations this imposes on high-frequency generalisation, certain strategies have been proposed such as dynamically increasing network capacity \cite{fang2024addressingspectralbiasdeep}, phase-shifted activation functions \cite{doi:10.1137/19M1310050} and adaptive Fourier bases \cite{JAGTAP2020109136}.

\textbf{Contributions} of this paper: Our work extends these studies by uniting the Fourier perspective of PQCs with spectral bias understood as a property of learning dynamics, explicitly relating loss gradients to the rate at which different frequency components of target functions are learned. We first establish a theoretical link between spectral bias and the redundancy structure of Fourier coefficients present in PQCs, showing that frequency components with high redundancy can exhibit larger gradients than those with lesser redundancy and do so on average when parameters are small. We then verify this numerically by comparing the learning dynamics across various encoding schemes and assess the robustness of each encoding strategy to perturbations of the trainable parameters. We empirically test the effects of parameter initialisation and entanglement structure on spectral training dynamics.

\section{Background}
\begin{figure}
\centering
\begin{quantikz}
    & \qwbundle{n} & \gate{W(\theta)} & \gate{S(x)}\gategroup[1,steps=2,style={inner
    sep=6pt}]{repeat $\times L$} & \gate{W(\theta)} & \meter{}
\end{quantikz}
\caption{General reuploader circuit design with trainable gates $W(\theta)$ and data encoding gates $S(x)$.}
\label{circ:basic_circuit}
\end{figure}
We will consider circuits of the general form seen in Figure \ref{circ:basic_circuit}, known as reuploader circuits, and we denote the function this circuit outputs as $f$. Circuits are defined on $n$ qubits with data encoding unitaries $S(x)$, trainable unitaries $W(\theta)$ with trainable parameters $\theta \in \Theta$ and a Hermitian observable $O$. This framework considers classical data $\mathcal{X}=(\vec{x}^{1}, \dots, \vec{x}^{m})$, along with encoding unitaries which encode each element $x_k$ of $\vec{x}$ onto one qubit via a gate $G(x_k)=e^{-i\beta x_kH_k}$, where $\beta$ is some scale factor one may wish to apply. The full encoding unitary is given by $S(x)=\prod_{k=1}^{n} e^{-i\beta_k x_kH_k}$. combining all unitaries into a single unitary $U(x,\theta)$ of dimension $2^n$, the output of the circuit then takes the form
\begin{equation}
    f_{\theta} = \langle 0 | U^{\dag}(x, \theta) O U(x, \theta) | 0 \rangle,    
\end{equation}

whereby $U(x,\theta)$ is defined as
\begin{equation}
    U(x, \theta) = \prod_{l=1}^{L}\left(W^l(\theta_{l})S^l(x)\right) W^{0}(\theta_{0}),
\end{equation}

where $L$ is the total number of reuploading layers in the circuit. By considering the construction above we know the circuit can be functionally represented as a Fourier series \cite{PhysRevA.103.032430}:
\begin{equation}
    f(x,\theta) = \sum_{\vec{\omega} \in \Omega} c_{\vec{\omega}}(\theta)e^{i\vec{\omega} \cdot \vec{x}}.
\label{eq:n_dim_fourier}
\end{equation}
For simplicity, we focus on a one-dimensional input case: 
\begin{equation}
    f(x) = \sum_{\omega\in \Omega} c_{\omega}e^{i\omega x}.
    \label{eq:circuit_output}
\end{equation}
The set of frequencies $\Omega$ to which the model can have access is determined by the eigenvalues of the encoding unitaries which can be assumed to be diagonal $S^l(x) = diag(\lambda^l_1, ..., \lambda^l_d)$. More specifically:
\begin{equation}
    \Omega = \{\omega = \Lambda\mathbf{_k} - \Lambda_\mathbf{j}, \mathbf{k}, \mathbf{j} \in [d]^L\}.
\end{equation}
Here, the multi-index notation  $\mathbf{j}=\{j_1, ..., j_L\}\in[d]^L$, has been introduced, where $[d]^L$ denotes the set of $L$ integers between $1,...,d$ and $d=2^n$, with $n$ denoting the number of qubits. The sum of eigenvalues coming from $L$ embedding gates can then be stated as $\Lambda\mathbf{_j}=\lambda_{j_1}+...+\lambda_{j_L}$ for a given $\mathbf{j}$. The coefficients $c_{\vec{\omega}}$ are generally nontrivial and depend on the non-embedding layers of the circuit and the observable $O$, which can also be assumed diagonal~\cite{Casas_2023}:
\begin{equation}\label{eq:freq_coeff}
    c_\omega = \sum_{\substack{\mathbf{k},\mathbf{j}\in[d]^L \\ \Lambda_k - \Lambda_j=\omega}} a_{\mathbf{k},\mathbf{j}},
\end{equation} with
\begin{equation}\label{eq:individual_coeff}
    a_{\mathbf{k},\mathbf{j}} = \sum_{i}{\left(O_i{W}_{k_L, i}^{*(L)}{W}_{i,j_L}^{(L)}\right)}{W}_{1,k}^{*(0)}{W}_{j_1,1}^{(0)}\prod_{p=2}^{L}{{W}_{k_{p-1},k_p}^{*(p-1)}W_{j_p,j_{p-1}}^{(p-1)}}.
\end{equation}

It is important to note multiple combinations of $\mathbf{k}, \mathbf{j}$ lead to identical values of $\Lambda\mathbf{_k} - \Lambda_\mathbf{j}$, which we shall refer to as the redundancy $R(\omega)$ of a frequency $\omega$.

\section{Spectral bias}\label{sec:spectral_bias}
In this section we present theoretical results about gradients of PQCs in regression tasks, based on their frequency spectra. 

Let us begin with stating the upper bounds of these gradients in two scenarios. First, Theorem~\ref{thm:upper_bound_int} describes a simple upper bound for models whose spectra consist of integer-valued frequencies. This is true for the most basic and proliferated version of the reuploader model~\cite{PhysRevA.103.032430}. In this scenario, the terms in Equation~\ref{eq:circuit_output} form an orthogonal basis for the space of square-integrable functions on $[0,2\pi]$:
\begin{equation}
    \int_{-\pi}^{\pi}e^{-ikx}e^{ijx} = 0 \Rightarrow j\neq k \quad j,k \in \mathbb{Z}.
\end{equation}

\begin{theorem}
    \textbf{Upper bound on the gradient of the loss at frequency $\omega$ for integer-frequency models}
    
    Let $f(x,\theta)$ denote the output of a PQC with an integer-valued spectrum, trained to minimise the mean squared loss with respect to a target function $h(x)$, both of which can be expressed as Fourier series with the same, integer-valued, spectrum. Under gradient descent, frequencies $\omega$ with larger redundancies $R(\omega)$ can induce larger gradients in the loss. Specifically, for any parameter $\theta$, the magnitude of the gradient of the loss at frequency $\omega$ satisfies:
    \begin{equation}
     \left|\partial_{\theta} L(\omega) \right| \leq 4R(\omega)||O||_{\text{tr}}\left|c_{D_\omega}\right|,
        \label{eq:upper_bound_int}
    \end{equation}
    where $||O||_{\text{tr}}$ is the trace norm of $O$, which, for Hermitian matrices, is the sum of absolute values of their eigenvalues. $c_{D_\omega}$ indicates the difference between the coefficients $c_\omega$ of the model and the target.
    \label{thm:upper_bound_int}
\end{theorem}

Theorem~\ref{thm:upper_bound_nonint} is an upper bound for a more general case, where the spectrum $\Omega$ contains arbitrary frequencies. This scenario is closer to what occurs, in models with trainable embeddings~\cite{Jaderberg_2024}. We note, however, that the theorem does not cover that case fully, as the target spectrum is assumed to match exactly that of the model. We consider the understanding of training dynamics of trainable-spectra models to be an interesting and important research direction.

\begin{theorem}
    \label{thm:upper_bound_nonint}
    \textbf{Upper bound on the gradient of the loss at frequency $\omega$ for arbitrary-frequency models}

    Let $f_{\cancel{{\perp}}}(x,\theta)$ denote the output of a PQC with a non-integer spectrum, trained to minimise the mean squared loss with respect to a target function $h(x)_{\cancel{{\perp}}}$, whose spectrum matches that of $f_{\cancel{{\perp}}}(x,\theta)$. Under gradient descent, frequencies $\omega$ induce gradients boosted by other frequencies $\omega'$, based on their proximity. Frequencies with larger redundancies $R(\omega)$ can contribute larger values to the gradients of the loss. Specifically, for any parameter $\theta$, the magnitude of the gradient of the loss at frequency $\omega$ satisfies:
\begin{equation}
    \begin{split}
    &|\partial_\theta L(\omega)| 
    \leq 2||O||_{\text{tr}}\sum_{\omega'}|\text{sinc}(\pi(\omega-\omega'))| \times \left( \left| c^*_{D_{\omega'}} \right| R(\omega) + \left| c_{D_\omega} \right| R(\omega') \right).
\end{split}
\end{equation}
\end{theorem}

To analyse this expression, we begin by noting the behaviour of the $\text{sinc(x)}$ function, which acts as a weighing term in the sum. Immediately, we see that the cross-term contributions will be suppressed for frequencies far from $\omega$. In the local neighbourhood of $\omega$, once again, frequencies with the largest redundancies can achieve largest gradients. For well-spread-out frequencies ($|\omega - \omega'| \gg 1$), the cross-terms disappear and we recover the result for orthogonal frequencies (Equation~\ref{eq:upper_bound_int}).

This shows that the gradient at a given frequency is allowed to be greater when $R(\omega)$ is greater. With the upper bounds for frequency-component gradients established, we now turn to finding their values in expectation.

The following theorems provide expected values of the gradients at each frequency, under the assumption of trainable single-qubit non-encoding unitaries with small parameters. Such an assumption holds, for example, for small-angle initialisation schemes \cite{PhysRevApplied.22.054005, zhang2025escapingbarrenplateaugaussian}. For this approach we use the formalism of PQCs provided in \cite{wiedmann2024fourieranalysisvariationalquantum}, whereby the circuits in consideration must consist of non-trainable Clifford gates and single-qubit Pauli rotations. The model can then be described as:
\begin{multline}
\label{eq:trig_monomial}
f(\boldsymbol\theta ,x)
=
\sum_{\substack{s,c\in\mathbb N_0^{d}\\ s',c'\in\mathbb N_0^{w}}}
k_{s,c,s',c'}\ 2^{-\sum_{j=1}^d(s_j + c_j)}(-i)^{\sum_{j=1}^{d}s_j}\\
\Bigg(\prod_{k=1}^{w}\sin^{s'_k}(\theta_k)\cos^{c'_k}(\theta_k)\Bigg)
\Bigg(\prod_{j=1}^d\sum_{a_j=0}^{s_j}\sum_{b_j=0}^{c_j}
\binom{s_j}{a_j}\binom{c_j}{b_j}(-1)^{\,s_j-a_j}\,
e^{i(2a_j+2b_j-s_j-c_j)x_j}\Bigg),
\end{multline}

The sum over nodes $(s, c, s', c')$, each of which are non-negative integer vectors $s,c\in\mathbb{N}^d_0, s',c'\in\mathbb{N}^w_0$. The variables $s, c$ denote the number of sine and cosine terms for each input $x_j$ and $s',c'$ likewise but for each variational parameter $\theta_k$. The parameters $k_{s,c,s',c'}$ are either $0$ or $1$ in absolute value, originating from expectation values. The trainable variational parameters are given by $\theta_k$ and the data $x_j = \beta_j x$, where $\beta_j$ is some constant. This is a slight modification from the original formulation in \cite{wiedmann2024fourieranalysisvariationalquantum} which considered $d$-dimensional data, instead we consider one-dimensional data but multiplied by some coefficient $\beta_j$. The frequencies of the Fourier decomposition are then given by 
\begin{equation}
\omega(\mathbf{a}, \mathbf{b}) = \sum_{j=1}^d m_j(a_j, b_j)\beta_{j} = \sum_{j=1}^d \omega_j .
\end{equation}

$\mathbf{a}=(a_1,\dots,a_d)$ and $\mathbf{b}=(b_1,\dots,b_d)$ are multi-indices describing which element of the double sum at each index $j$ was used to create a given frequency. The total frequency $\omega$ associated with the specific choice $(\mathbf{a}, \mathbf{b})$ is the sum of the local frequency contributions determined by $a_j$ and $b_j$ and $m_j(a_j, b_j) := 2a_j+2b_j - s_j - c_j \in \mathbb Z$. The Fourier coefficients are then given by 
\begin{equation}
\label{eq:trig_coeff}
c_\omega(\boldsymbol\theta)
=
\sum_{s,c,s',c'} k_{s,c,s',c'}\ 2^{-\sum_{j=1}^d(s_j + c_j)}(-i)^{\sum_{j=1}^{d}s_j}\;p(s, c, \omega)\;
\prod_{k=1}^w \sin^{s'_k}(\theta_k)\cos^{c'_k}(\theta_k).
\end{equation}
where 
\begin{equation}
p(s, c, \omega) = \prod_{j=1}^d\sum_{a_j=0}^{s_j}\sum_{b_j=0}^{c_j} \left[ \binom{s_j}{a_j}\binom{c_j}{b_j}(-1)^{s_j-a_j} \delta^{\omega_j}_{m_j(a_j,b_j)\beta_j} \right].
\end{equation}

allows us to group the contributions to a given frequency $\omega$. The overall model spectrum is the set of frequencies that survive possible cancellations across nodes (different nodes with the same variational polynomial can cancel for all $\theta$ only if their total prefactor vanishes). The final spectrum is then
\[
\Omega = \left\{ \omega |\
\exists\ s',c'\ \text{ such that }\
\sum_{s,c} k_{s,c,s',c'}\,p(s, c, \omega)\neq 0
\right\}.
\]

The redundancy $R(\omega)$ is defined as the number of distinct variational terms contributing to the frequency $\omega$. It corresponds to the number of non-vanishing summands in the expression for $c_\omega(\boldsymbol\theta)$. Formally, it is the cardinality of the set of active multi-indices:
\[
R(\omega) := \Big| \Big\{ (s,c,s',c') \;\Big|\; k_{s,c,s',c'}\, p(s,c,\omega) \neq 0 \Big\} \Big|.
\]
This value $R(\omega)$ determines the number of components summing to form the coefficient.

\begin{theorem}    \label{thm:small_angle1}
\textbf{Expected upper bound on the gradient of the loss at frequency $\omega$ for integer frequencies.}

Under the assumption models are initialized with small-angles from a Gaussian distribution $\mu=0, \sigma\ll1$. The magnitude of the gradient of the loss at frequency $\omega$ to first-order satisfies:
\begin{equation}
    \mathbb{E}\left[ \left| \partial_{\theta_k} L(\omega) \right| \right] \lesssim 2 |c_{\omega_h}| \sum_{r=1}^{R(\omega)} \frac{|p(\mathbf{s}^{(r)},\mathbf{c}^{(r)},\omega)|}{2^{\sum^d_{j=1}(s^{(r)}_j+c^{(r)}_j)}} \left[ \frac{s'^{(r)}_k 2^{\frac{s'^{(r)}_k-1}{2}}\Gamma(\frac{s'^{(r)}_k}{2})}{\sqrt{\pi}} \prod_{j\neq k} \frac{2^{\frac{s'^{(r)}_j}{2}}\Gamma(\frac{s'^{(r)}_j+1}{2})}{\sqrt{\pi}} \right] \sigma^{S^{(r)}-1}.
\end{equation}
$\Gamma$ functions represent the absolute moments of Gaussian random variables, $S(r)$ is the total sine-degree from summing over all paths. In the case of no weight-sharing thus $s'^{(r)}_j$=1 at most.
\begin{equation}
    \mathbb{E}\left[ \left| \partial_{\theta_k} L(\omega) \right| \right] \lesssim 2 |c_{\omega_h}| \sum_{r=1}^{R(\omega)} \frac{|p(\mathbf{s}^{(r)},\mathbf{c}^{(r)},\omega)|}{2^{\sum (s^{(r)}_j+c^{(r)}_j)}} \left( \frac{2}{\pi} \right)^{\frac{w-1}{2}} \sigma^{S^{(r)}-1}.
\end{equation}

Demonstrating Fourier coefficients with high redundancy can potentially elicit stronger gradient signals in expectation. Refer to Appendix~\ref{appx:exp_upper_bound_int} for the full proof.

\end{theorem}

We can similarly, demonstrate this for the case of non-integer frequencies.

\begin{theorem}{Expected upper bound on the gradient of the loss at frequency $\omega$ for non-integer frequencies.}

\begin{multline}\mathbb{E}[|\partial_{\theta_k} L(\omega)|] \lesssim \sum_{\omega'} |\text{sinc}(\pi(\omega-\omega'))| \Bigg( \|h_{\omega'}| \sum_{r=1}^{R(\omega)} \mathcal{G}^{(r)}(\omega) \sigma^{S^{(r)}-1}+ |h_{\omega}| \sum_{q=1}^{R(\omega')} \mathcal{G}^{(q)}(\omega') \sigma^{S^{(q)}-1} \Bigg),\end{multline}

where $\mathcal{G}^{(r)}(\omega)$ represents the explicit geometric and combinatorial prefactor for the $r$-th path of frequency $\omega$:
\begin{equation}\mathcal{G}^{(r)}(\omega) = \frac{|p(\mathbf{s}^{(r)},\mathbf{c}^{(r)},\omega)|}{2^{\sum(s^{(r)}_j+c^{(r)}_j)}} \left[ \frac{s'^{(r)}_k 2^{\frac{s'^{(r)}_k-1}{2}}\Gamma(\frac{s'^{(r)}k}{2})}{\sqrt{\pi}} \prod_{j\neq k} \frac{2^{\frac{s'^{(r)}_j}{2}}\Gamma(\frac{s'^{(r)}_j+1}{2})}{\sqrt{\pi}} \right].
\end{equation}

Refer to Appendix~\ref{appx:exp_upper_bound_nonint} for a full proof.

\end{theorem}

While the preceding results establish upper bounds, the expected upper bounds derived here further emphasise the role of redundancy in spectral bias, providing sharper guarantees on average. Furthermore, redundancy analysis allows us to highlight that terms with high redundancy can be expected to exhibit higher gradients. Thus, one can use the redundancy distribution of an encoding scheme to boost gradients of frequencies of interest. On the other hand, the Heisenberg evolution picture used in the derivation shows a suppression of terms which pick up many trigonometric terms. Highest frequencies are always formed from such paths, indicating the inevitable tail-off of any encoding scheme.

\section{Experimental results}\label{sec:spectral_results}
\begin{figure}
\centering
\scalebox{0.6}{
\begin{quantikz}
    & \gate{R_y(\theta_{000})} & \gate{R_x(\theta_{001})} & \ctrl{1} & & \gate{R_x(\beta_0 x)}\gategroup[5,steps=5,style={inner
        sep=6pt}]{repeat $\times L$} & \gate{R_y(\theta_{0i0})} & \gate{R_x(\theta_{0i1})} & \ctrl{1} & &\meter{} \\
    & \gate{R_y(\theta_{100})} & \gate{R_x(\theta_{101})}& 
    \targ{} & \ctrl{1} & \gate{R_x(\beta_1 x)} & \gate{R_y(\theta_{1i0})} & \gate{R_x(\theta_{1i1})} & \targ{} & \ctrl{1} &\\
    & \gate[nwires=0,style={fill=white,draw=white,text height=1cm}]{\vdots} & \gate[nwires=0,style={fill=white,draw=white,text height=1cm}]{\vdots}& \gate[nwires=0,style={fill=white,draw=white,text height=1cm}]{\vdots} \ctrl{1}& \gate[nwires=0,style={fill=white,draw=white,text height=1cm}]{\vdots}
    & \gate[nwires=0,style={fill=white,draw=white,text height=1cm}]{\vdots} & \gate[nwires=0,style={fill=white,draw=white,text height=1cm}]{\vdots} & \gate[nwires=0,style={fill=white,draw=white,text height=1cm}]{\vdots} & \gate[nwires=0,style={fill=white,draw=white,text height=1cm}]{\vdots} \ctrl{1}& \gate[nwires=0,style={fill=white,draw=white,text height=1cm}]{\vdots} &\\
    & \gate{R_y(\theta_{n-200})} & \gate{R_x(\theta_{n-201})} & \targ{} & \ctrl{1} & \gate{R_x(\beta_{n-2} x)} & \gate{R_y(\theta_{n-2i0})} & \gate{R_x(\theta_{n-2i1})}& \targ{} & \ctrl{1} &\\
    & \gate{R_y(\theta_{n-100})} & \gate{R_x(\theta_{n-101})} & & \targ{} & \gate{R_x(\beta_{n-1}x)} & \gate{R_y(\theta_{n-1i0})} & \gate{R_x(\theta_{n-1i1})} && \targ{} &
\end{quantikz}
}
\caption{The specific reuploader model used for the experiments, containing trainable parameters $\theta$ and data $x$. The choice of coefficients $\beta$ determines the nature of the encoding and $L$ is the number of circuit layers.}
\label{circ:spectral_bias_circuit}
\end{figure}

\begin{figure}[hb!]
    \centering
    \begin{subfigure}[b]{0.49\linewidth}
        \includegraphics[width=\linewidth]{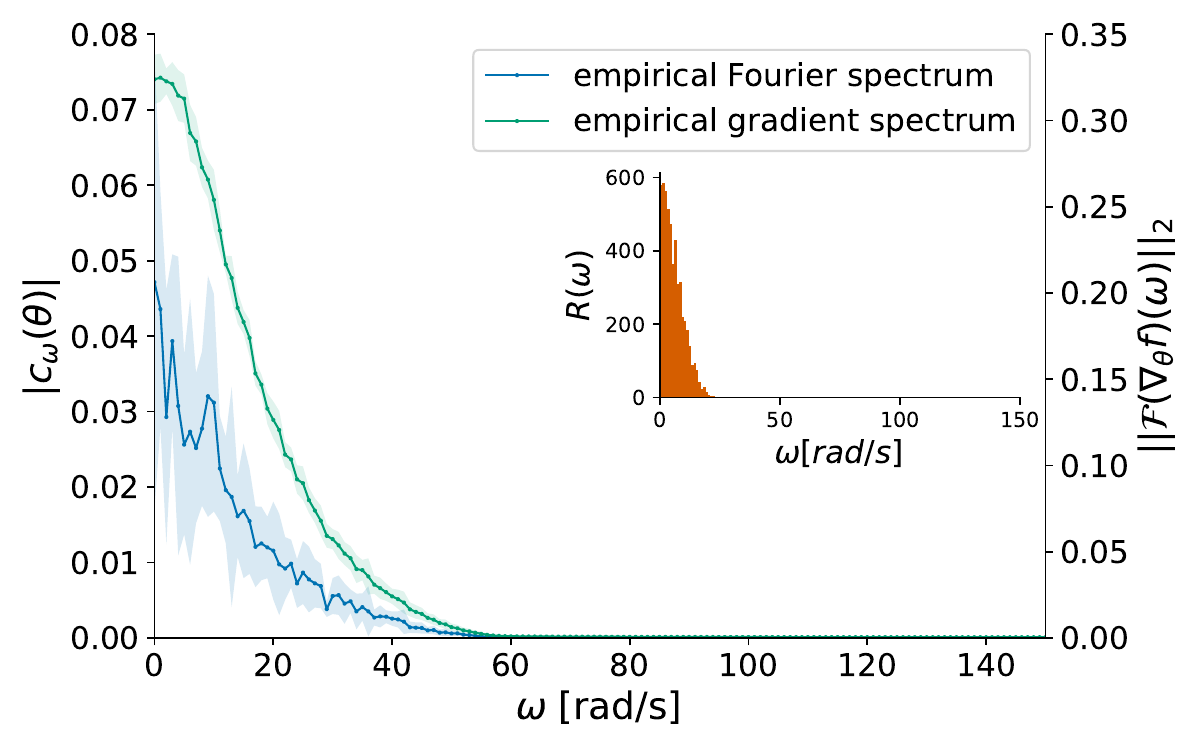}
        \caption{Constant Pauli encoding}
        \label{fig:redundancy_a}
    \end{subfigure}
    \hfill
    \begin{subfigure}[b]{0.49\linewidth}
        \includegraphics[width=\linewidth]{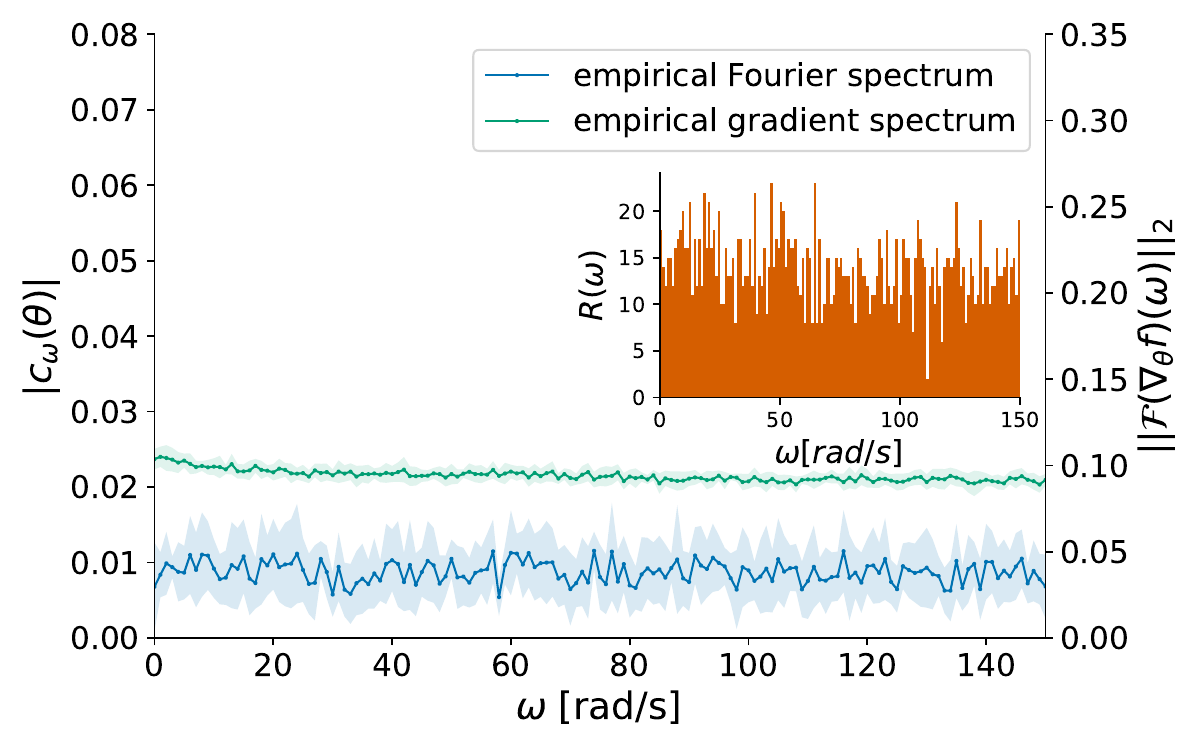}
        \caption{Ternary Pauli encoding}
        \label{fig:redundancy_b}
    \end{subfigure}

    \caption{The Fourier spectra of the two Pauli encoding schemes constant and ternary with empirical data taken as the sampled mean over ten models. Depicted are the sampled theoretically accessible frequencies (light red), the mean Fourier coefficient (blue), and the total gradient of trainable parameters at each Fourier coefficient (green).}
    \label{fig:redundancy}
\end{figure}
Models used in this section follow the general structure of Fig.~\ref{circ:spectral_bias_circuit} with $5$ qubits, $L=20$ and $O = Z_1$. The following results presented follow a similar analysis to that presented in the work \cite{cao2020understandingspectralbiasdeep}. We perform a regression task with each of the models, introducing a target function of the form 
\begin{equation}
    h(x) = \sum_{\omega \in \Omega_h} A_\omega \sin(\omega x + \phi_\omega),
\end{equation}
with frequencies $\omega$ forming the set $ \Omega_h = \{5, 10, 15, 20, 25, 30, 35, 40, 45, 50\}$, all $A_\omega$ set to $1$ before normalization and the phases $\phi_\omega$ drawn from the uniform distribution over $U(0,2\pi)$ for each training instance. We sample $2048$ equally spaced points in $x$ in the interval $x\in[0,2\pi]$. 
We train the models using \texttt{PyTorch} \cite{Ansel_PyTorch_2_Faster_2024} and \texttt{Pennylane} \cite{bergholm2022pennylaneautomaticdifferentiationhybrid} using the Adam optimizer \cite{kingma2017adammethodstochasticoptimization}, evaluating every $5$ epochs. The batch size is equal to the train size and the learning rate is fixed at $0.0005$. All results reported are statistical means computed from ten separate runs. Compute resources can be found in Appendix \ref{appx:compute}.
\subsection{Spectral bias}
\label{subsec:exp_spectral_bias}
The main goal of our work is to show that spectral bias in PQCs is tightly related to the redundancy of a given frequency in a model. To do so, we introduce two models (two more models are shown in the Appendices), each with a different encoding scheme. The models differ only by their embedding coefficients $\beta_i$ (See Figure~\ref{circ:spectral_bias_circuit}). We refer to the first type of embedding as the constant Pauli embedding, where all $\beta_i=1$, which results in a concentrated frequency spectrum. The other model considered is the ternary exponential Pauli embedding of \cite{PhysRevA.107.012422}, with coefficients $\beta_i=3^{i}$. This encoding admits a much more uniform spectrum. In fact, it generates the widest possible spectrum per reupload layer for single-qubit embedding gates. In our experiments, the coefficients $\beta_i$ are repeated in each of the $L$ reupload layers. One could use a scheme in which the powers of $3$ continue to rise with each reupload layer, which would result in a wider and flatter spectrum. We note however, that such coefficients would quickly become numbers too large to process efficiently.

Before training these models, we depict their frequency spectrum in Figure \ref{fig:redundancy}, to aid the above argument. What we observe is that the coefficients in the models as well as the gradient spectrum follow a trend based on the redundancy spectrum of the model. Results for a wider number of encodings can be found in Appendix~\ref{appx:redundancy_analysis_extended}.

\begin{figure}[t!]
    \centering
    \begin{subfigure}[b]{0.45\linewidth}
        \includegraphics[width=\linewidth]{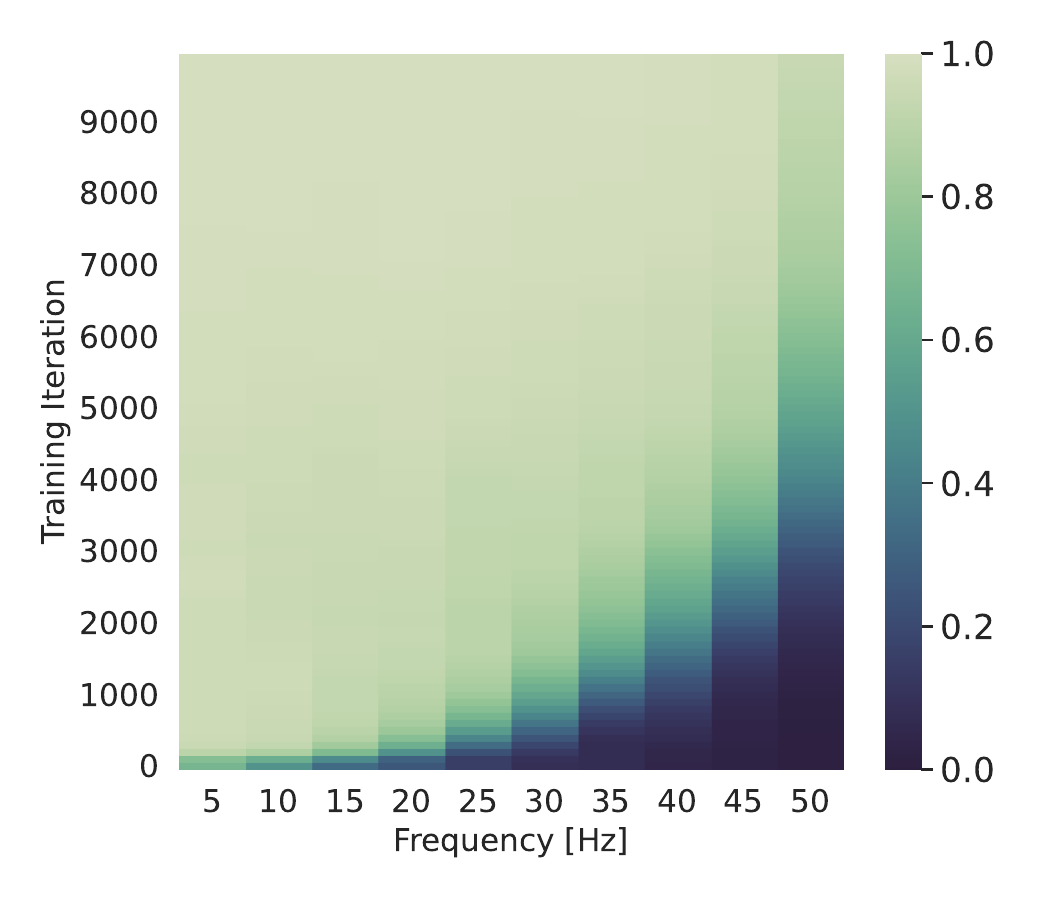}
        \caption{Constant encoding}
    \label{fig:spectral_bias_a}
    \end{subfigure}
    \begin{subfigure}[b]{0.45\linewidth}
        \includegraphics[width=\linewidth]{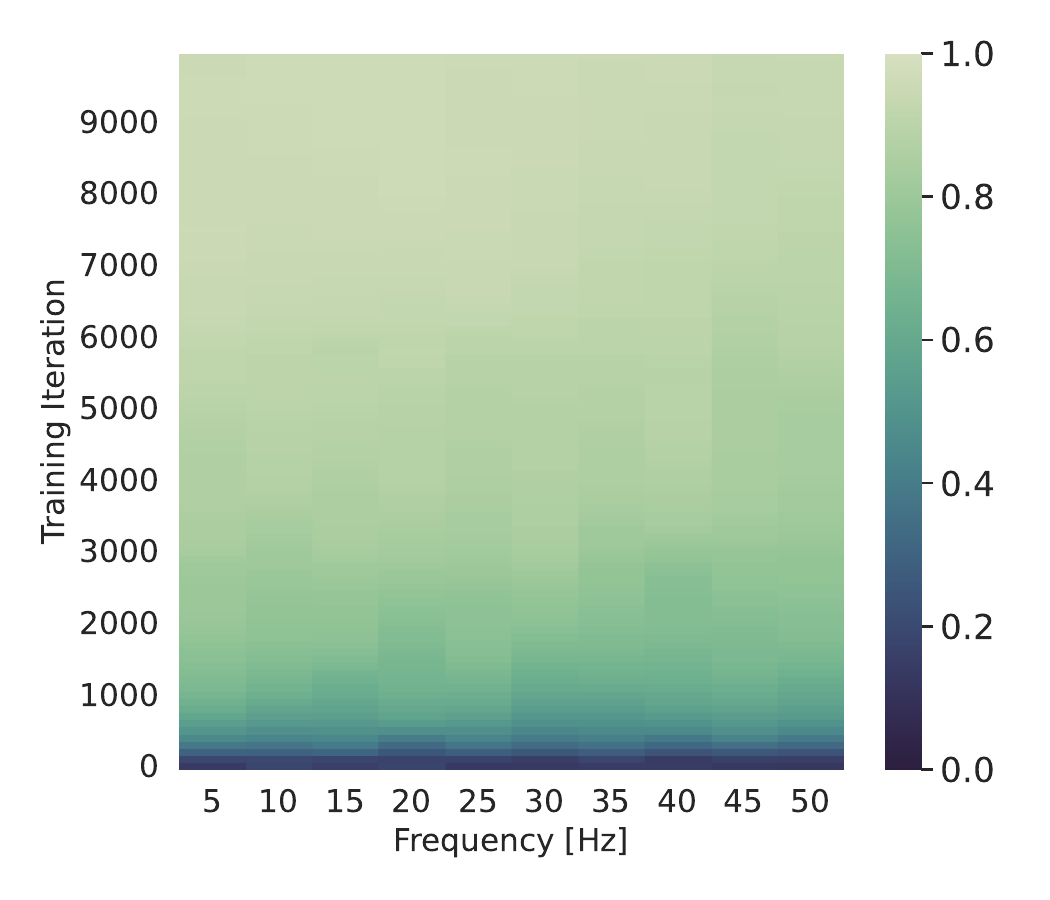}
        \caption{Ternary exponential encoding}
    \label{fig:spectral_bias_b}
    \end{subfigure}

    \caption{The rate at which frequencies (x-axis) are learnt during the course of training (y-axis), the colorbar measures the PQC spectrum normalised by the target amplitude at a given frequency ($|\tilde{f}_\omega|/A_i$). Each subplot depicts the training dynamics for a different encoding scheme.}
    \label{fig:spectral_bias}
\end{figure}

The results of fitting the target function can be seen in Figure \ref{fig:spectral_bias}. The emerging trend is clear; frequencies with higher redundancy are learnt faster. Specifically, the constant Pauli encoding, where redundancy falls off sharply, takes a long time to learn high frequencies and the exponential encoding, where redundancy is roughly constant, learns all frequencies at equal rates. Refer to Appendix~\ref{appx:spectral_bias_extended}, for a similar analysis on a greater number of encodings.

\subsection{Robustness}

We continue our exploration of the properties of PQCs by evaluating the robustness of the models' parameters to perturbations. Let $c_\omega(\theta^*)$ be the learned Fourier coefficient at frequency $\omega$, and suppose it decomposes as a sum of $R(\omega)$ contributing terms $c_\omega(\theta^*) \;=\; \sum_{i=1}^{R(\omega)} a_i(\theta^*)$. We then can study the robustness of the models under small isotropic parameter perturbations.

\begin{theorem}[Upper bound on the robustness of PQCs to isotropic parameter perturbations]\label{thm:robustness} The normalized root mean square (RMS) deviation is upper-bounded by
\begin{equation}\label{eq:general-bound}
\mathcal{R} \;=\; \frac{\sqrt{\mathbb{E}[|\Delta C|^2]}}{|c(\theta^*)|}
\;\le\; \frac{\sigma_a}{\kappa\,\bar a}\,\sqrt{\frac{1+(R(\omega)-1)\rho}{R(\omega)}}.
\end{equation}

where $\sigma_a$ is the typical perturbation scale of each summand and $\bar a$ is a typical per-summand magnitude, $\kappa$ the coherence between terms $a_i$ and $\rho$ the correlation between terms $a_i$. The full proof can be found in Appendix~\ref{appx:robustness_proof}
\end{theorem}

In particular, if \(\rho\approx 0\) (perturbation effects approximately uncorrelated across \(i)\)) and \(\kappa\) is bounded away from zero, then
\[
\mathcal{R} \lesssim \frac{1}{\kappa}\,\frac{\sigma_a}{\bar a}\,\frac{1}{\sqrt{R(\omega)}}.
\]

Thus, in the case where per-term perturbations are approximately uncorrelated and the summands add with non-negligible coherence, redundancy reduces relative sensitivity roughly as $R(\omega)^{-1/2}$.  If instead the perturbation responses are strongly correlated or the summands cancel in phase, the redundancy benefit vanishes.

Continuing with the setup from section \ref{subsec:exp_spectral_bias}, we evaluate each model at the end of training and examine the effect of random isotropic perturbations $\vec{\theta} = \vec{\theta^*} + \delta \hat{\theta}$ to the learnt function. Here, $\delta$ is a chosen magnitude and $\hat{\theta}$ is a random unit vector in parameter space. Once $\vec{\theta}^*$ is perturbed we compute $f$ at the new parameters $\vec{\theta}$ and find the coefficients of the model at the frequencies of interest from the target function. Averages are taken over $100$ samples of $\hat{\theta}$. A final average is taken over the phases $\phi$ of the target function.

\begin{figure}[]
    \centering
    \begin{subfigure}[b]{0.49\linewidth}
        \includegraphics[width=\linewidth]{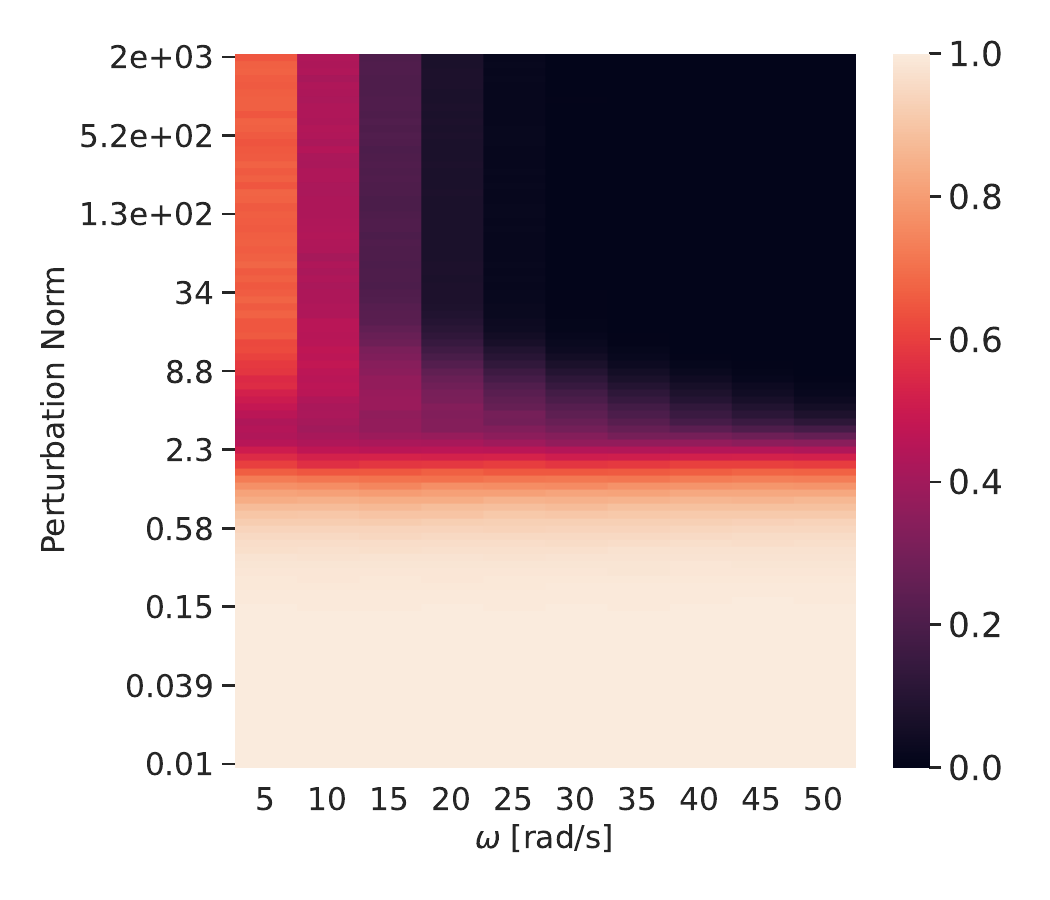}
        \caption{Constant Pauli encoding}
    \label{fig:robustness_a}
    \end{subfigure}
    \begin{subfigure}[b]{0.49\linewidth}
        \includegraphics[width=\linewidth]{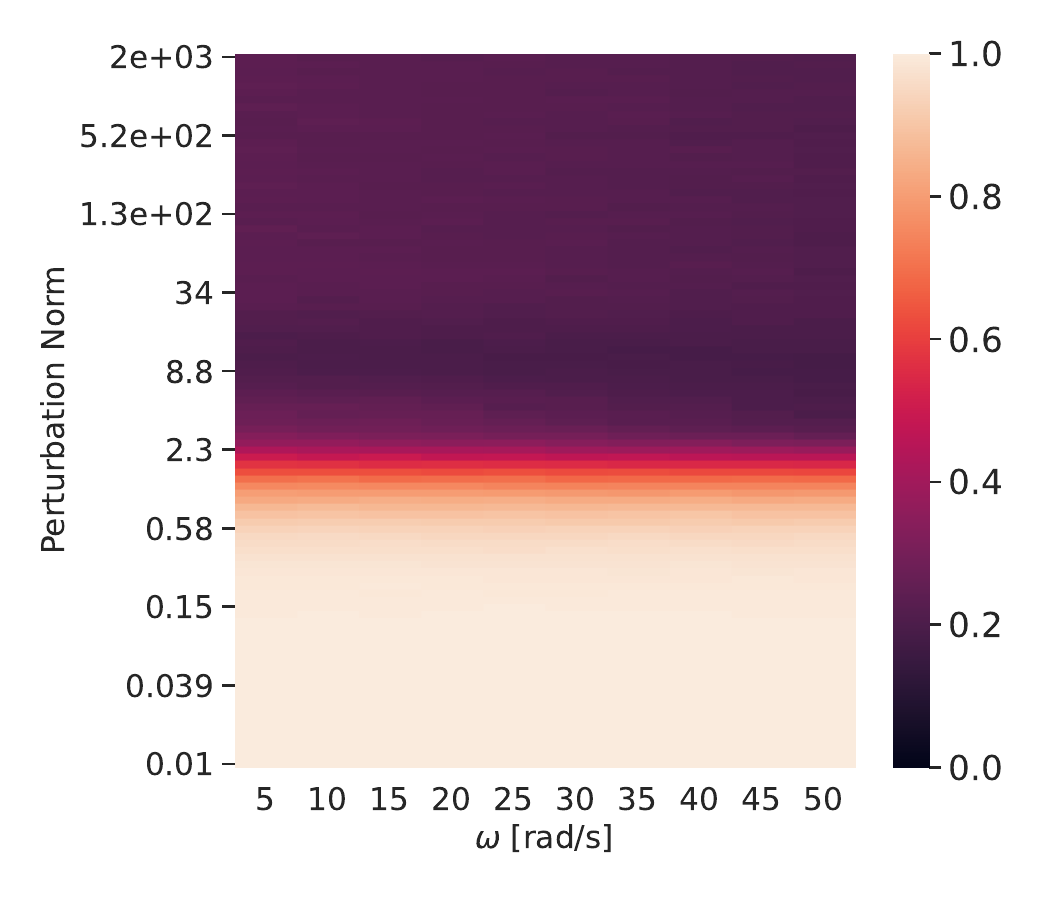}
        \caption{Ternary Pauli encoding}
    \label{fig:robustness_b}
    \end{subfigure}

    \caption{Normalised Fourier spectrum of the model output (x-axis: frequency, colourbar: magnitude) as a function of parameter perturbation (y-axis). Each subplot depicts the effects of parameter perturbations for a different encoding scheme.}
    \label{fig:robustness}
\end{figure}

Figure~\ref{fig:robustness} depicts the results for two encoding schemes with their trainable final parameters perturbed. It is evident lower frequencies are more robust to perturbations for the constant Pauli model whereas the exponential Pauli model shows a uniformity to the robustness at each frequency. We can see here the circuits operate in the regime where redundancy plays a key role in the robustness of the model at each frequency.

\subsection{Entanglement}
\begin{figure}[t!]
    \centering
        \begin{subfigure}[b]{\linewidth}
        \includegraphics[width=\linewidth]{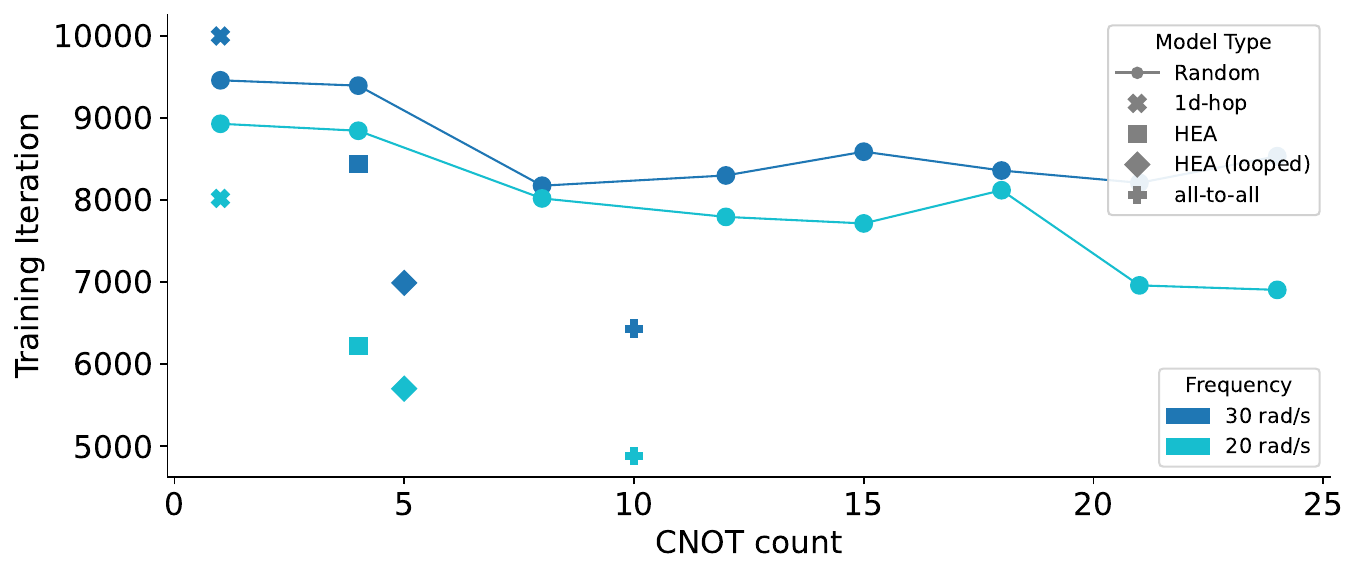}
        \caption{}\label{fig:entanglement_a}

    \end{subfigure}
    \begin{subfigure}[b]{0.49\linewidth}
        \includegraphics[width=\linewidth]{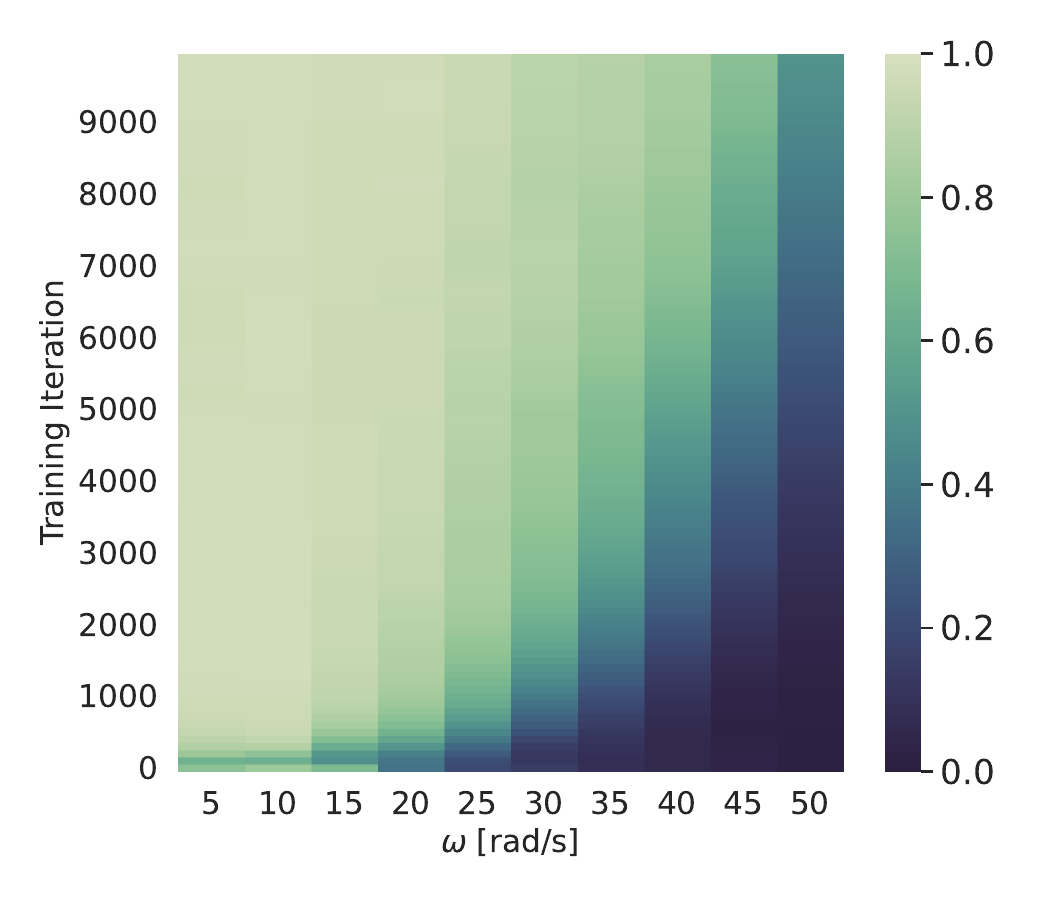}
        \caption{1d-hop entanglement}\label{fig:entanglement_b}
    
    \end{subfigure}
    \hfill
    \begin{subfigure}[b]{0.49\linewidth}
        \includegraphics[width=\linewidth]{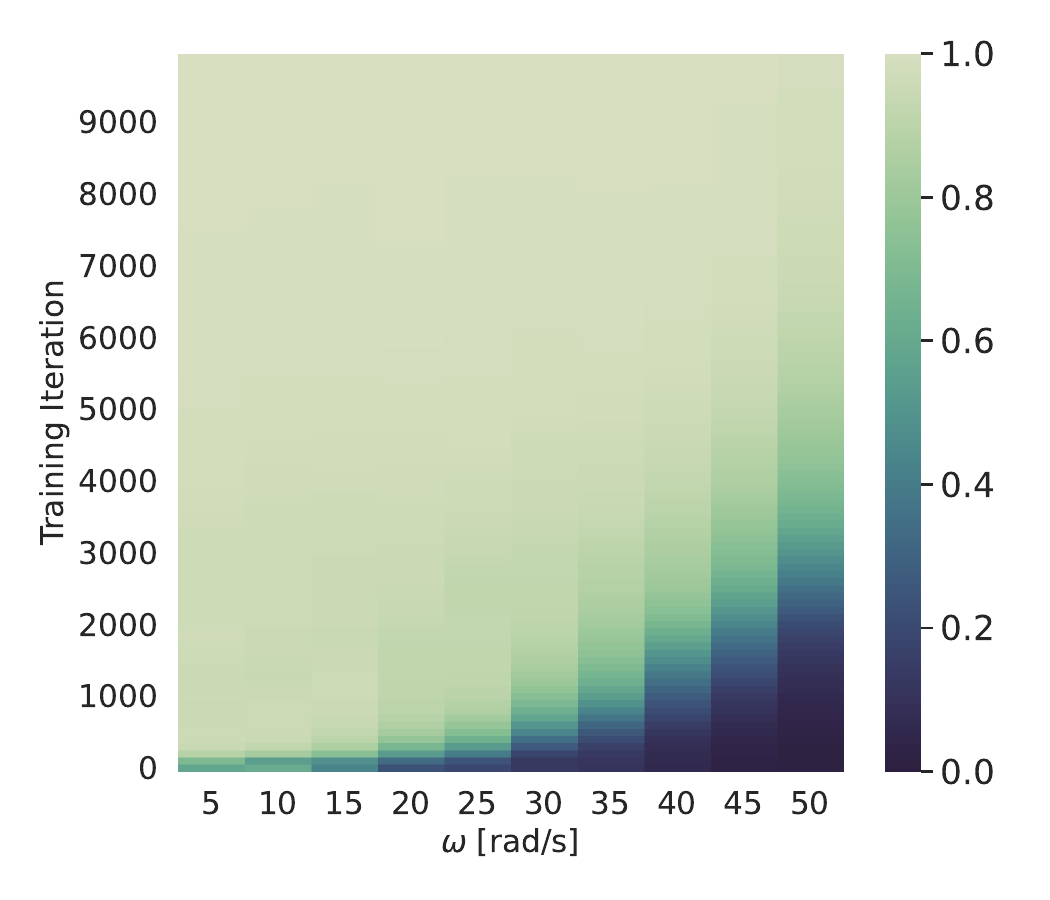}
        \caption{all-to-all entanglement}\label{fig:entanglement_c}
    
    \end{subfigure}

    \caption{The effect of entanglement on spectral training. (a) shows the number of epochs needed for a model to learn a frequency based on the number of CNOT gates randomly placed in each layer of the ansatz. The y-axis denoting the mean number of epochs until convergence to a target frequency amplitude, and the x-axis denotes the number of CNOT gates in a single layer of the circuit. The solid line represents the number of epochs until convergence for random CNOT placement and the scattered points are more structured entanglement structures. (b) and (c) depict the spectral dynamics for two of the entanglement structures 1d-hop and all-to-all respectively}
    \label{fig:entanglement}
\end{figure}

In this section we examine the effect of the entanglement structure on the spectral dynamics, the results of which are in Figure~\ref{fig:entanglement}. It is known that parallel encoding provides a model with more frequencies. This is only true, however, if the additional qubits are entangled (either in the circuit or via a multi-qubit measurement). It is of interest then, to probe how entanglement affects the resulting spectral bias. 
We use the constant Pauli encoding scheme and vary the entangling scheme. For the experiment we select four well known entangling structures and compare the number of epochs each takes to converge to a given amplitude associated to a frequency component of the target. We also, include a baseline of randomly placed CNOT gates. The structured models have their convergence values taken from a mean over ten random initialisations, while the randomised models are means over 20. The final results are in Figure~\ref{fig:entanglement_a}, which shows that an increase in the number of CNOT reduces the effect of spectral bias, even if the CNOT gates are randomly placed, though a more structured placement yields a greater reduction. We also highlight two of the structured ansatz spectral dynamics. The first model, places one nearest-neighbour entanglement gate per reuploading layer, in a cascading way. The Second, connects all qubits to one each other. Both can be seen in Figures~\ref{fig:entanglement_b} and~\ref{fig:entanglement_c} respectively, depicting their spectral dynamics. It is clear that higher degrees of entanglement reduce the spectral bias. A theoretical understanding of this effect, combined with the insights about redundancy discovered in this work could provide a powerful toolkit for designing circuits with custom spectral bias, tailored for a given task. The reduction in spectral bias from the addition of CNOTs can be reasoned from a lightcone argument starting from the measurement operator, whereby the increase in CNOT gates reduces the number of excluded parameters contributing to that given operators expectation value. 

\subsection{Initialization}

In this section, we investigate how the scale of parameter initialization affects the rate at which different frequencies are learned in PQCs. We restrict our study to circuits using constant Pauli encoding and initialise the trainable parameters $\theta$ from a normal distribution $\mathcal{N}(0, \sigma^2)$ where the variance $\sigma^2$ is varied across experiments.

\begin{figure}[t!]
    \centering
    \includegraphics[width=\linewidth]{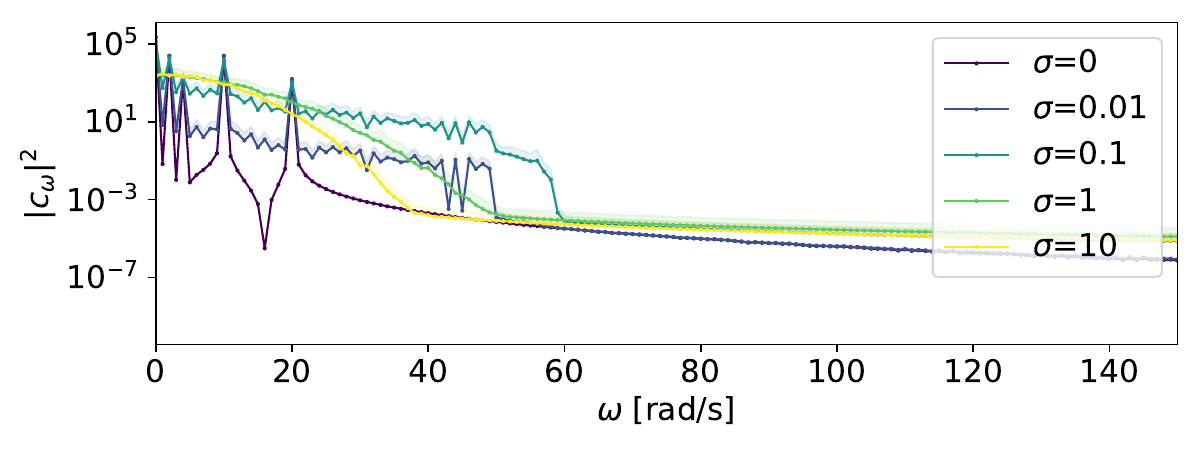}
    \caption{Size of Fourier coefficients based on the variance (around $0$) of the trainable PQC parameters.}
    \label{fig:init_coeffs}
\end{figure}

Before training, we examine the spectrum of PQC model outputs under different initialisation scales by computing the Fourier coefficients of the circuits output. In Figure \ref{fig:init_coeffs}, we plot the squared magnitudes of the coefficients for increasing values of the intialisation standard deviation, from $0.01$ to $10$. We see that as $\sigma$ increases from zero $|c_{\omega}|^2$ initially increases until $0.1$ before decreasing in magnitude across all frequencies. This implies small initializations tend to initialise circuits with larger Fourier coefficients than larger initialisations.

\begin{figure}[t!]
    \centering
    \begin{subfigure}[b]{0.4\linewidth}
        \includegraphics[width=\linewidth]{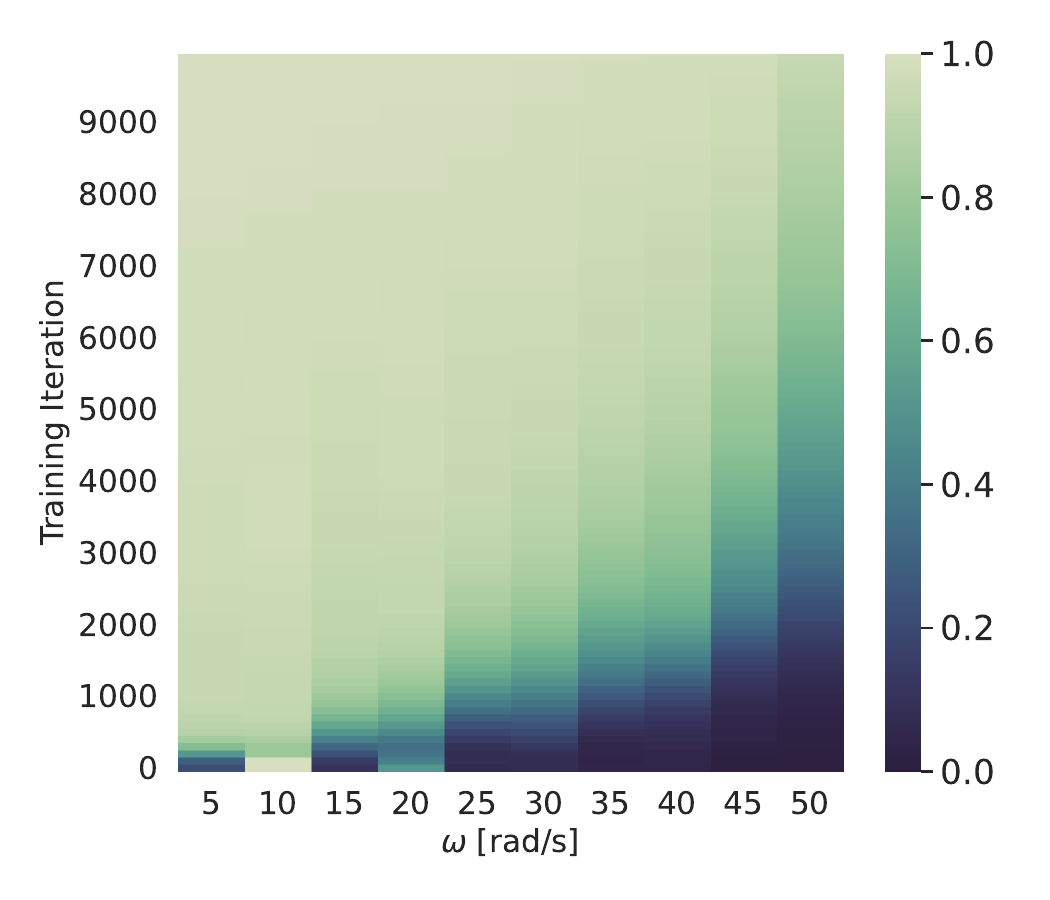}
        \caption{$\sigma^2=0.01$}
    \label{fig:init_a}
    \end{subfigure}
    \begin{subfigure}[b]{0.4\linewidth}
        \includegraphics[width=\linewidth]{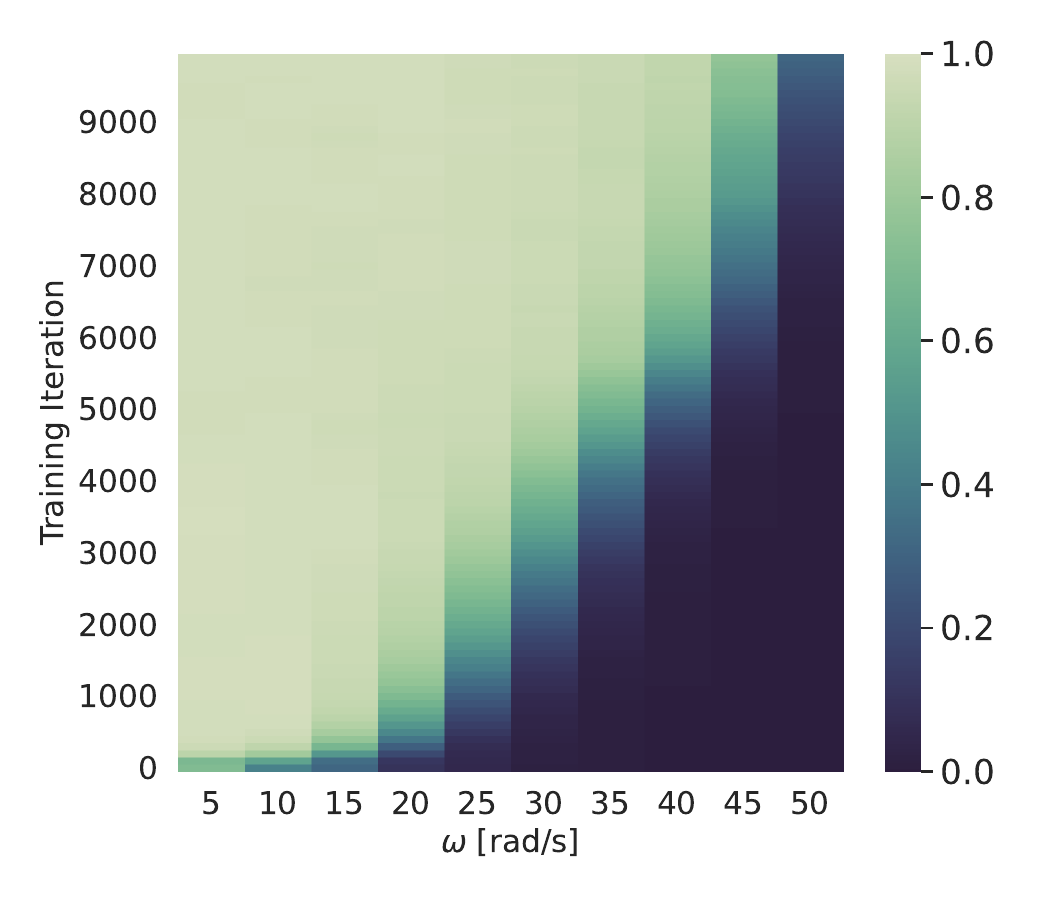}
        \caption{$\sigma^2=10$}
    \label{fig:init_b}
    \end{subfigure}
    \caption{The rate at which frequencies (x-axis) are learnt during the course of training (y-axis), the colorbar measures the PQC spectrum normalised by the target amplitude at a given frequency ($|\hat{f}_\omega|/A_i$). Each subplot depicts the training dynamics for a different initialization scheme (a) $\sigma^2=0.01$, (b) $\sigma^2=10$.}
    \label{fig:init}
\end{figure}

This suppression of coefficients has practical implications. In Figure \ref{fig:init}, we show the spectral dynamics of training under two initialisation regimes. When initialised with $\sigma=0.01$ in Figure \ref{fig:init_a} the circuit begins with relatively large coefficient magnitudes across many frequencies, and the spectral bias is modest. However, when initialized with $\sigma=10$ in Figure \ref{fig:init_b}, the Fourier coefficients particularly those at higher frequencies start off significantly smaller, leading to smaller gradient magnitudes and slower learning. As a result, the spectral bias is exacerbated.

In summary, the initialisation scale not only determines the expressivity at the start of training but also fundamentally shapes the learning dynamics across the frequency spectrum. Care should be taken as to how one initialises a PQC especially for tasks involving high-frequency content. Future work should look into formalising the observed decrease in Fourier coefficient amplitude as the size of initialization is increased.

\section{Conclusion}
In this work, we have provided the first rigorous investigation into the spectral bias of PQCs, grounding our analysis in their Fourier structure. We established a theoretical link between the training dynamics of PQCs and the redundancy of Fourier coefficients, defined as the number of terms contributing to a given frequency component. The magnitude of the gradient of the loss with respect to a circuit parameter at a particular frequency is shown to be upper bounded by the frequency's redundancy. This provides a mechanism by which PQCs can exhibit spectral bias. Through numerical experiments, we confirmed this theoretical relationship across multiple encoding schemes, demonstrating that models with encoding strategies that distribute redundancy uniformly (exponential Pauli encoding) learn frequencies at equal rates. Conversely, encodings with steep redundancy decay, like constant Pauli encoding, show markedly slower convergence at higher frequencies. We further explored how model design affects spectral learning dynamics. We found that increasing the variance of parameter initialisation suppresses the initial magnitude of Fourier coefficients across all frequencies, exacerbating spectral bias and slowing learning, particularly high frequency components. The entangling scheme also affects spectral bias in a similar fashion with lesser entanglement contributing to the effect. Frequency redundancy was also seen to aid model robustness, whereby under global parameter perturbations greater redundnacy led to greater robustness at these frequencies. Despite these insights, our study has several limitations. Most notably, we focus on single-qubit encodings with integer frequencies and synthetic problems. Future work may aim to generalise our analysis to a broader class of PQCs, including those with parameter sharing, mid-circuit measurements and complex encodings along with practical problems of interest. Moreover, finding theoretical results for the robustness, entangling and parameter initalization results would be fruitful avenues to pursue. Ultimately, understanding spectral bias in PQCs will shed light on where these model may find use in the real world.
\FloatBarrier
\medskip
\bibliographystyle{plainnat}
\bibliography{references}

\appendix
\section{Compute Resources}\label{appx:compute}
All experiments were conducted using two eight-core Intel Xeon E5-2620 3.0GHz CPU along with 64GB of RAM and 16GB of storage for the results.

\section{Proof of Theorem~\ref{thm:upper_bound_int}}
\begin{proof}
    Assuming the target function $h(x)$ lies in the family of models expressible by $f(x,\theta)$ (and thus described by Equation~\ref{eq:circuit_output}), define the difference between the model output and the target
\begin{equation}
    D(x) = \left|h(x) - f(x)\right| = \sum^{N}_{\omega=-N}c_{D_\omega}e^{i\omega x},
\end{equation} 
where $c_{D_\omega}$ is the difference between the Fourier coefficients $c_\omega$ of $f(x,\theta)$ and $h(x)$ respectively. The mean-squared loss over the domain of $x \in [0, 2\pi]$ is thus
\begin{equation}
    L = \frac{1}{2\pi}\int_{0}^{2\pi} D(x)^2 dx.
\end{equation}
By Parseval's theorem \cite{Stein2003FourierAA}, this is simply equal to:
\begin{equation}
    L = \sum^{N}_{\omega=-N}\left|c_{D_\omega}\right|^2 = \sum^{N}_{\omega=-N}L(\omega).
\label{eq:freq_loss}
\end{equation}

As a result, $L(\omega) = |c_{D_\omega}|^2$. The gradient of the loss with respect to a circuit parameter $\theta$ is then
\begin{equation}
\partial_\theta L= \sum_{\omega=-N}^{N}\partial_\theta L(\omega)
\label{eq:freq_loss_deriv}
\end{equation}

To ascertain whether a spectral bias exists within PQCs we focus on the terms in the summand of equation \ref{eq:freq_loss_deriv} to find the contribution from an $\omega$ at $\theta$, to identify whether gradients at certain values of $\omega$ are larger than others. Each term satisfies

\begin{equation}
    \partial_\theta L(\omega)= c_{D_\omega}(\theta)\partial_\theta c^*_{D_\omega}(\theta)+ c^*_{D_\omega}(\theta)\partial_\theta c_{D_\omega(\theta)}.
\end{equation}

Since the coefficients of $h(x)$ have no dependence on $\theta$ their derivatives are zero. What remains are derivatives for the Fourier coefficients of $f(x,\theta)$ which we shall denote $c_{\omega_f}$:
\begin{equation}
    \partial_\theta L(\omega) = c_{D_\omega}(\theta)\partial_\theta c^*_{\omega_f}(\theta)+ c^*_{D_\omega}(\theta)\partial_\theta c_{\omega_f}(\theta),
\end{equation}

The magnitude of the gradient is bounded as follows,
\begin{equation}
    \left| \partial_\theta L(\omega) \right| = \left | 2\text{Re}\left(c^*_{D_\omega}(\theta)\partial_\theta c_{\omega_f}(\theta)\right)\right| \leq 2\left|c_{D_\omega}(\theta)\right|\left|\partial_\theta c_{\omega_f}(\theta)\right|.
\end{equation}
Which depends, intuitively, on the difference between the target and model coefficients, and the derivative of the model's coefficient itself. The derivative of $c_{\omega_f}$ can be further upper-bounded from its decomposition:
\begin{equation}
    \left|\partial_\theta c_\omega\right| \leq \sum_{\substack{\mathbf{k},\mathbf{j}\in[d]^L \\ \Lambda_k - \Lambda_j=\omega}} \left |\partial_\theta a_{\mathbf{k},\mathbf{j}}(\theta)\right|.
\end{equation}
$a_{\mathbf{k},\mathbf{j}}$ can be seen as a weighted sum over the eigenvalues of the Hermitian observable $O$, with weights formed by products of unitary matrix entries from the parameterised gates. Since any given $\theta$ is present only in a single layer, it can be shown that 
\begin{equation}
\label{eq:freq_coeff}
\left|\partial_\theta a_{\mathbf{k},\mathbf{j}}(\theta) \right| \leq 2||O||_{\text{tr}}.
\end{equation}
Altogether we find
\begin{equation}
     \left|\partial_\theta L(\omega)\right| \leq 4R(\omega)||O||_{\text{tr}}\left|c_{D_\omega}(\theta)\right|.
     \label{eq:appx_upper_bound_int}
\end{equation}
\label{proof:up_bound_int}
\end{proof}

\section{Proof of Theorem~\ref{thm:upper_bound_nonint}}
\begin{proof}
    \label{proof:up_bound_nonint}
Following the analysis for integer-frequency models (Proof~\ref{proof:up_bound_int}), we study the squared error loss in the frequency domain:

\begin{equation}
\begin{split}
    L &= \frac{1}{2\pi}\int_{0}^{2\pi}(f_{\cancel{\perp}}(x)-h(x))^2dx \\
    &= \frac{1}{2\pi}\int_{0}^{2\pi}(\sum_{\omega\in \Omega_{\cancel{\perp}}}{c_{D_\omega}e^{i\omega x}})^2dx \\
    &= \frac{1}{2\pi}\int_{0}^{2\pi}\sum_{\omega}\sum_{\omega'}c_{D_\omega} c^*_{D_{\omega'}} e^{ix(\omega-\omega')}dx \\
    &=\sum_{\omega}\sum_{\omega'}c_{D_\omega} c^*_{D_{\omega'}}e^{i\pi(\omega-\omega')}\text{sinc}(\pi (\omega - \omega')),
\end{split}
\end{equation}

with cross-terms ($\omega \neq \omega'$). There is now an ambiguity in assigning part of the total sum loss to a certain frequency $L(\omega)$, which was not present in the integer-only spectrum models. Without loss of generality, we choose to assign $L(\omega) = \text{Re}\left(c_{\omega}\sum_{\omega'}c^*_{\omega'}e^{i\pi(\omega-\omega')}\text{sinc}(\pi (\omega - \omega'))\right)$.
To determine an upper bound for $\left|\partial_\theta L(\omega)\right|$, let us first denote $\pi(\omega - \omega') = \beta_{\omega'}$ and thus:
\begin{equation}
    \left|\partial_\theta L(\omega)\right| 
    =
    \left |\partial_\theta\text{Re}\left(\sum_{\omega'}e^{i\beta_{\omega'}}\text{sinc}(\beta_{\omega'})c_{D_\omega} c^*_{D_{\omega'}} \right ) \right |
    =
    \left |\text{Re}\left(\sum_{\omega'}e^{i\beta_{\omega'}}\text{sinc}(\beta_{\omega'})\partial_\theta \left(c_{D_\omega} c^*_{D_{\omega'}}\right )\right)  \right |
\end{equation} 
and using properties of the absolute value:
\begin{equation}
    \left|\partial_\theta L(\omega)\right| 
   \leq 
    \left |\sum_{\omega'}e^{i\beta_{\omega'}}\text{sinc}(\beta_{\omega'})\partial_\theta \left(c_{D_\omega} c^*_{D_{\omega'}}\right )  \right |
    \leq 
    \sum_{\omega'}|\text{sinc}(\beta_{\omega'})| \times\left | \partial_\theta \left (c_{D_\omega} c^*_{D_{\omega'}}\right)\right | .
\end{equation}
The derivative in a single element in this sum satisfies:
\begin{equation}
   \left|\partial_\theta(c_{D_\omega} c^*_{D_{\omega'}}) \right| 
   =
   \left|c^*_{D_{\omega'}} \partial_\theta c_{f_\omega} +  c_{D_\omega} \partial_\theta c^*_{f{\omega'}}\right| 
   \leq 
   \left|\partial_\theta c_{f_\omega} \right| \left| c^*_{D_{\omega'}} \right| + \left| c_{D_\omega} \right| \left| \partial_\theta c^*_{f_{\omega'}}\right|,
\end{equation}
and using $\left|\partial_\theta c_{\omega} \right| \leq 2R(\omega) ||O||_{\text{tr}}$, established in Proof~\ref{proof:up_bound_int}, 
\begin{equation}
    \left|\partial_\theta(c_{D_\omega} c^*_{D_{\omega'}}) \right| 
    \leq 
    2||O||_{\text{tr}}\left( \left| c^*_{D_{\omega'}} \right| R(\omega) + \left| c_{D_\omega} \right| R(\omega') \right).
\end{equation}
Which results in 
\begin{equation}
    \begin{split}
    &|\partial_\theta L(\omega)| 
    \leq 2||O||_{\text{tr}}\sum_{\omega'}|\text{sinc}(\pi(\omega-\omega'))| \times \left( \left| c^*_{D_{\omega'}} \right| R(\omega) + \left| c_{D_\omega} \right| R(\omega') \right)
\end{split}
\end{equation}
\end{proof}

\section{Small angle approximation}

\subsection{Expected value}\label{sec:exp_val}
We analyse the expected gradient magnitude of the Fourier coefficients under a small-angle initialization of the variational parameters. Assuming each parameter is drawn independently from a Gaussian distribution $\theta \;\overset{\text{iid}}{\sim}\; \mathcal{N}(0,\sigma^2)$ with $\sigma \ll 1$, our goal is to compute (maybe add second too?)
\begin{equation}
    \mathbb{E}[\,|\partial_{\theta_k} c_\omega(\boldsymbol{\theta})|\,],    
\end{equation}

which is the expected absolute gradient of a Fourier coefficient with respect to a single parameter $\theta_k$. We factorise the monomial representation of $c_\omega$ and focus initially only on the terms dependent on trainable parameters $\boldsymbol{\theta}$ and let
\begin{equation}
    g(\boldsymbol{\theta}) = \prod_{j=1}^{w} g_j(\theta_j), 
    \qquad g_j(\theta_j) = \sin(\theta_j)^{s'_j}\cos(\theta_j)^{c'_j},
\end{equation}
where $s'_j, c'_j \in \mathbb{Z}_{\ge 0}$. Fixing an index $k$, taking the partial derivative with respect to $\theta_k$ isolates the $k$-th term
\begin{equation}
    \frac{\partial g}{\partial \theta_k} 
    = \partial_{\theta_k} g_k(\theta_k)\prod_{j\neq k} g_j(\theta_j),
\end{equation}
with
\begin{equation}
    \partial_{\theta_k}g_k(\theta_k) = s'_k\,\sin^{s'_k-1}(\theta_k)\cos^{c'_k+1}(\theta_k)
    - c'_k\,\sin^{s'_k+1}(\theta_k)\cos^{c'_k-1}(\theta_k).
\end{equation}

By independence of the $\theta_j$ and multiplicativity of the absolute value,
\begin{equation}
\mathbb{E}\!\left[\,\Big|\frac{\partial g}{\partial \theta_k}\Big|\,\right] 
    = \mathbb{E}[\,|g'_k(\theta_k)|\,] \prod_{j\neq k}\mathbb{E}[\,|g_j(\theta_j)|\,].
\end{equation}
We now begin with applying the small-angle approximation, firstly by writing $\partial_{\theta_k}g_k(\theta_k) = A(\theta_k) - B(\theta_k)$ with
\begin{align}
    A(\theta_k) &= s'_k\,\sin^{s'_k-1}(\theta_k)\cos^{c'_k+1}(\theta_k),\\
    B(\theta_k) &= c'_k\,\sin^{s'_k+1}(\theta_k)\cos^{c'_k-1}(\theta_k).
\end{align}

Inserting the expansions of $\sin$ and $\cos$ around zero:
\begin{align}
    \sin^r(\theta) &= \theta^r\left(1 - \frac{r}{6}\theta^2 + O(\theta^4)\right),\\
    \cos^t(\theta) &= 1 - \frac{t}{2}\theta^2 + O(\theta^4),
\end{align}

Applying these approximations to the undifferentiated monomials $g_j(\theta_j) = \sin^{s'_j}(\theta_j)\cos^{c'_j}(\theta_j)$ yields the local expansion:
\begin{equation}
\label{eq:g_j}
    g_j(\theta_j) = \theta_j^{s'_j}\left(1 - \left(\frac{s'_j}{6} + \frac{c'_j}{2}\right)\theta_j^2 + O(\theta_j^4)\right).
\end{equation}

For the differentiated term $\partial_{\theta_k} g_k$, we proceed by substituting the trigonometric expansions into the components $A(\theta_k)$ and $B(\theta_k)$ defined above:
\begin{align}
    A(\theta_k) &= s'_k\,\theta^{s'_k-1}\Big(1 - \tfrac{s'_k-1}{6}\theta_k^2 + O(\theta_k^4)\Big)
                \Big(1 - \tfrac{c'_k+1}{2}\theta_k^2 + O(\theta_k^4)\Big) \\
              &= s'_k\,\theta^{s'_k-1}\left(1 - \Big(\tfrac{s'_k-1}{6} + \tfrac{c'_k+1}{2}\Big)\theta_k^2 + O(\theta_k^4)\right),\\
    B(\theta_k) &= c'_k\,\theta^{s'_k+1}\Big(1 - \Big(\tfrac{s'_k+1}{6} + \tfrac{c'_k-1}{2}\Big)\theta_k^2 + O(\theta_k^4)\Big).
\end{align}
Subtracting gives
\begin{equation}
\label{eq:dg_k}    \partial_{\theta_k}g_k(\theta) = s'_k\,\theta^{s'_k-1} - \frac{(s'_k+3c'_k)(s'_k+2)}{6}\,\theta^{s'_k+1} + O(\theta^{s'_k+3}).
\end{equation}
Factorising out the leading term and taking the absolute value,
\begin{equation}
    |\partial_{\theta_k}g_k(\theta)| = s'_k|\theta|^{s'_k-1}\left(1 - \frac{(s'_k+3c'_k)(s'_k+2)}{6s'_k}\,\theta_k^2 + O(\theta_k^4)\right).
\end{equation}
This assumes the term in the bracket is positive which requires
\begin{equation}
    |\theta_k| \lesssim \sqrt{\frac{6s'_k}{(s'_k+3c'_k)(s'_k+2)}}, \quad s'_k\geq 1.    
\end{equation}
Under regimes with no weight sharing ($s'_k,c'_k\leq1$), the tightest constraint occurs at $s'_k=1, c'_k=0$, requiring $|\theta_k| \lesssim 1/\sqrt{2} \approx 0.71$. This condition is satisfied with high probability for Gaussian initialization with $\sigma \ll 1$. For the case $s'_k=0$, the sine power vanishes and the expansion simplifies to
\begin{equation}
    \partial_{\theta_k}g_k(\theta_k) = -c'_k\,\sin(\theta_k)\cos^{c'_k-1}(\theta_k) = -c'_k\theta_k + O(\theta_k^3),
\end{equation}
yielding the magnitude
\begin{equation}
   \label{eq:dg_k_2} |\partial_{\theta_k}g_k(\theta_k)| \approx c'_k\,|\theta_k| + O(\theta_k^3).
\end{equation}

The absolute moments of Gaussian variables for $\theta \;\overset{\text{iid}}{\sim}\; \mathcal{N}(0,\sigma^2)$ are \cite{papoulis2002probability}
\begin{equation}
    \label{eq:moments}
    M_r := \mathbb{E}[|\theta|^r] 
    = \sigma^r 2^{r/2}\frac{\Gamma\!\left(\tfrac{r+1}{2}\right)}{\sqrt{\pi}}.
\end{equation}
Along with the recurrence relation
\begin{equation}
    \frac{M_{r+2}}{M_r} = \sigma^2(r+1).
\end{equation}

We can then use these moments to find the expectation value of $\mathbb{E}
[\,|\partial_{\theta_k}g(\theta_k)|\,]$ for $s'_k\ge 1$:
\begin{equation}
    \mathbb{E}[\,|\partial_{\theta_k}g_k(\theta_k)|\,] 
    = s'_k M_{s'_k-1} \left(1 - \frac{(s'_k+3c'_k)(s'_k+2)}{6 }\sigma^2 + O(\sigma^4)\right)
    \label{eq:exp_diff_gk}
\end{equation}

For $s'_k=0$,
\begin{equation}
    \mathbb{E}[\,|\partial_{\theta_k}g_k(\theta_k)|\,] = c'_k\,\mathbb{E}[|\theta_k|] + O(\sigma^3)
    = c'_k\,\sigma\sqrt{\tfrac{2}{\pi}} + O(\sigma^3).
\end{equation}
Now we can combine this, with the full expression for $g(\theta)$, whereby the expectations of $g_j(\theta_j)$ follows very similarly from above (maybe also state expansion above). For $s'_k\ge 1$, combining Equation~\ref{eq:exp_diff_gk} with the expectation value of Equation~\ref{eq:g_j} gives:
\begin{multline}
    \mathbb{E}[\,|\partial_{\theta_k}g(\boldsymbol{\theta})|\,] 
    = s'_kM_{s'_{k-1}}
      \left(1 - \tfrac{(s'_k+3c'_k)(s'_k+2)}{6}\sigma^2 + O(\sigma^4)\right) \\
      \times \prod_{j\neq k}\left(M_{s_j}
      \left(1 - \Big(\tfrac{s_j}{6} + \tfrac{c_j}{2}\Big)(s_j+1)\sigma^2 + O(\sigma^4)\right)\right).
\end{multline}

and for $s'_k=0$
\begin{equation}
    \mathbb{E}[\,|\partial_{\theta_k}g(\boldsymbol{\theta})|\,] 
    = \left( c'_k\,M_1 \right)
       \prod_{j\neq k}\left(M_{s_j}
      \left(1 - \Big(\tfrac{s_j}{6} + \tfrac{c_j}{2}\Big)(s_j + 1)\sigma^2 + O(\sigma^4)\right)\right).
\end{equation}

The Fourier coefficient $c_\omega$ (Equation~\ref{eq:trig_coeff})is a linear combination of such monomials, with prefactors $k_{s,c,s',c'}$, phases, and combinatorial factors $p(s,c,\omega)$, thus taking the expectation value its gradient:
\begin{equation}
    \mathbb{E}[\,|\partial_{\theta_k} c_\omega(\boldsymbol{\theta})|\,]
    \leq \sum_{s,c,s',c'} |k_{s,c,s',c'}| \;
      2^{-\sum_j(s_j+c_j)}|(-i)^{\sum_j s_j}|\,p(s,c,\omega)\;
      \mathbb{E}[\,|\partial_{\theta_k}g(\boldsymbol{\theta})|\,].
      \label{eq:exp_del_c}
\end{equation}

Likewise, $\mathbb{E}[c_{\omega}(\theta)]$:
\begin{equation}
    \mathbb{E}[\,| c_\omega(\boldsymbol{\theta})|\,]
    \leq \sum_{s,c,s',c'} |k_{s,c,s',c'}| \;
      2^{-\sum_j(s_j+c_j)}(-i)^{\sum_j s_j}\,p(s,c,\omega)\;
      \mathbb{E}[\,|g(\boldsymbol{\theta})|\,].
\end{equation}

Noting that the prefactors $|k_{s,c,s',c'}|$ are either $0$ or $1$, we rewrite the expectation in Eq.~\ref{eq:exp_del_c} by explicitly enumerating the $R(\omega)$ non-zero contributing variational paths. Let the index $r$ denote the $r$-th active path contributing to frequency $\omega$, with associated exponents $\mathbf{s}^{(r)}, \mathbf{c}^{(r)}, \mathbf{s}'^{(r)}, \mathbf{c}'^{(r)}$. For the dominant case ($s'_k \ge 1$), the expected gradient is bounded by the sum over these $R(\omega)$ components to leading order, explicitly subsituting in the moments from Equation~\ref{eq:moments}:
\begin{equation}
    \mathbb{E}[\,|\partial_{\theta_k} c_\omega(\boldsymbol{\theta})|\,] \lesssim \sum_{r=1}^{R(\omega)} \frac{|p(\mathbf{s}^{(r)},\mathbf{c}^{(r)},\omega)|}{2^{\sum(s^{(r)}_j+c^{(r)}_j)}} \left[ \frac{s'^{(r)}_k 2^{\frac{s'^{(r)}_k-1}{2}}\Gamma(\frac{s'^{(r)}_k}{2})}{\sqrt{\pi}} \prod^{w^{(r)}}_{j\neq k} \frac{2^{\frac{s'^{(r)}_j}{2}}\Gamma(\frac{s'^{(r)}_j+1}{2})}{\sqrt{\pi}} \right] \sigma^{S^{(r)}-1},
\end{equation}
and similarly for the suppressed case ($s'_k=0$):
\begin{equation}
    \mathbb{E}[\,|\partial_{\theta_k} c_\omega(\boldsymbol{\theta})|\,] \lesssim \sum_{r=1}^{R(\omega)} \frac{|p(\mathbf{s}^{(r)},\mathbf{c}^{(r)},\omega)|}{2^{\sum(s^{(r)}_j+c^{(r)}_j)}} \left[ c'^{(r)}_k \sqrt{\frac{2}{\pi}} \prod^{w^{(r)}}_{j\neq k} \frac{2^{\frac{s'^{(r)}_j}{2}}\Gamma(\frac{s'^{(r)}_j+1}{2})}{\sqrt{\pi}} \right] \sigma^{S^{(r)}+1}.
\end{equation}
Here $S^{(r)} = \sum_j s'^{(r)}_j$ is the total sine-degree of the $r$-th monomial. These expressions explicitly link the expected value of the gradient magnitude of frequency components to the redundancy $R(\omega)$, showing it is a coherent sum of $R(\omega)$ terms, each scaled by the initialization variance $\sigma$. For the case of no weight sharing and thus $s'_k,c'_k$ at most are one. To first order
\begin{equation} \mathbb{E}[\,|\partial_{\theta_k} c_\omega(\boldsymbol{\theta})|\,] \lesssim \sum_{r=1}^{R(\omega)} \frac{|p(\mathbf{s}^{(r)},\mathbf{c}^{(r)},\omega)|}{2^{\sum(s^{(r)}_j+c^{(r)}_j)}} \left[ \left( \frac{2}{\pi} \right)^{\frac{{w^{(r)}}-1}{2}} \right] \sigma^{S^{(r)}-1}. 
\end{equation}

\subsection{Second Moment}\label{sec:second_moment}

To compute the second moments, we require the moments of the squared Gaussian variable $\theta^2$ (where $\theta \overset{\textit{iid}}{\sim} \mathcal{N}(0, \sigma^2)$), defined as:
\begin{equation}
    \mu_{2n} := \mathbb{E}[\theta^{2n}] = (2n-1)!!\,\sigma^{2n}
\end{equation}
The recurrence relation for these moments is $\mu_{2n+2} = (2n+1)\sigma^2 \mu_{2n}$. We evaluate the expected squared gradient for the monomial $g(\boldsymbol{\theta})$:
\begin{equation}
    \mathbb{E}\left[\left(\frac{\partial g}{\partial \theta_k}\right)^2\right] 
    = \mathbb{E}\left[(\partial_{\theta_k} g_k(\theta_k))^2\right] \prod_{j\neq k} \mathbb{E}\left[g_j(\theta_j)^2\right],
\end{equation}

as well as $\mathbb{E}[g(\theta)]$.\newline

\textbf{Local Parameter Moments:} First we find the expectation values for the individual terms comprising $g(\theta)$.

\textit{Undifferentiated Terms ($g_j$):} Using the local expansion for $g_j(\theta_j)$ derived previously in \eqref{eq:g_j}, $g_j(\theta_j) \approx \theta_j^{s'_j}(1 - (s'_j/6 + c'_j/2)\theta_j^2)$, we square the expression and retain terms up to $\mathcal{O}(\theta_j^2)$ relative to the leading order:
\begin{equation}
    g_j(\theta_j)^2 \approx \theta_j^{2s'_j}\left(1 - \left(\frac{s'_j}{3} + c'_j\right)\theta_j^2\right).
\end{equation}
Taking the expectation yields:
\begin{align}
    \mathbb{E}[g_j(\theta_j)^2] &\approx \mu_{2s'_j} - \left(\frac{s'_j}{3} + c'_j\right)\mu_{2s'_j+2} \\
    &= (2s'_j-1)!!\,\sigma^{2s'_j} \left( 1 - \left(\frac{s'_j}{3} + c'_j\right)(2s'_j+1)\sigma^2 + \mathcal{O}(\sigma^4) \right).
\end{align}

\textit{Differentiated Term ($\partial_{\theta_k} g_k$):} We consider the two cases for the sine-degree  $s'_k$, $s'_k\geq1$ and $s'_k=0$.
\begin{itemize}
    \item \textit{Case ($s'_k \ge 1$):} The derivative scales with the sine power reduced by one. Squaring the expansion derived in Eq.~\eqref{eq:dg_k} retaining terms up to $\mathcal{O}(\theta_k^2)$ relative to leading order:
    \begin{align}
        \mathbb{E}[(\partial_{\theta_k} g_k)^2] &\approx (s'_k)^2 \mu_{2(s'_k-1)} - \frac{s'_k(s'_k+3c'_k)(s'_k+2)}{3} \mu_{2s'_k} \\
        &= (s'_k)^2 (2s'_k-3)!!\,\sigma^{2(s'_k-1)} \left( 1 - \frac{(s'_k+3c'_k)(s'_k+2)(2s'_k-1)}{3s'_k}\sigma^2 \right).
    \end{align}
    \item \textit{Case ($s'_k = 0$):} Squaring Eq.\eqref{eq:dg_k_2} gives $(\partial_{\theta_k}g_k)^2\approx(c'_k)^2\theta^2_k$, thus:
    \begin{equation}
        \mathbb{E}[(\partial_{\theta_k} g_k)^2] \approx (c'_k)^2 \sigma^2.
    \end{equation}
\end{itemize}

\textbf{Global Monomial Moments:} We now combine these local results to find the expected moments for the full monomial $g(\boldsymbol{\theta}) = \prod g_j(\theta_j)$.\newline

\textit{Undifferentiated Monomial Magnitude ($g(\mathbb{\theta})$):}
The expected squared magnitude of the monomial is the product of the undifferentiated expectations:
\begin{equation}
    \mathbb{E}[g(\boldsymbol{\theta})^2] = \prod_{j=1}^w \mathbb{E}[g_j(\theta_j)^2] \approx \left[ \prod_{j=1}^w (2s'_j-1)!! \right] \sigma^{2S}.
\end{equation}
where $S = \sum s'_j$ is the total sine-degree.

\textit{Monomial Gradient Magnitude:} The expected squared gradient combines the differentiated term $k$ with the product of undifferentiated terms $j \neq k$.
\begin{itemize}
    \item \textit{Case $s'_k \ge 1$:}
    \begin{multline}
        \mathbb{E}\left[\left(\partial_{\theta_k} g\right)^2\right] 
        = \left( (s'_k)^2 (2s'_k-3)!! \, \sigma^{2(s'_k-1)} \left[ 1 - \mathcal{O}(\sigma^2) \right] \right) \\
        \times \prod_{j\neq k} \left( (2s'_j-1)!! \, \sigma^{2s'_j} \left[ 1 - \mathcal{O}(\sigma^2) \right] \right).
    \end{multline}
    \item \textit{Case $s'_k = 0$:}
    \begin{equation}
         \mathbb{E}\left[\left(\partial_{\theta_k} g\right)^2\right] = \left( (c'_k)^2 \sigma^2 \right) \prod_{j\neq k} \left( (2s'_j-1)!! \, \sigma^{2s'_j} \left[ 1 - \mathcal{O}(\sigma^2) \right] \right).
    \end{equation}
\end{itemize}

\textbf{Fourier Coefficient:} The Fourier coefficient $c_\omega$ is a linear combination of these monomials. We will want to find the Root Mean Square (RMS) magnitudes. To bound the RMS amplitudes, we apply the triangle inequality for the $L_2$ norm ($\sqrt{\mathbb{E}[(\sum X)^2]} \le \sum \sqrt{\mathbb{E}[X^2]}$).\newline

\textit{RMS Coefficient Magnitude:} Summing the RMS of the monomials $\sqrt{\mathbb{E}[g^2]} \sim \sigma^S$:
\begin{equation}
    \sqrt{\mathbb{E}[|c_\omega(\boldsymbol{\theta})|^2]} \lesssim \sum_{r=1}^{R(\omega)} \frac{|p(\mathbf{s}^{(r)},\mathbf{c}^{(r)},\omega)|}{2^{\sum(s^{(r)}_j+c^{(r)}_j)}} \left[ \prod_{j} \sqrt{(2s'^{(r)}_j-1)!!} \right] \sigma^{S^{(r)}}.
\end{equation}

\textit{RMS Gradient Magnitude:}
Summing the RMS of the gradients $\sqrt{\mathbb{E}[(\partial g)^2]}$:
For the dominant case ($s'_k \ge 1$), we obtain:
\begin{equation}
    \sqrt{\mathbb{E}[|\partial_{\theta_k} c_\omega|^2]} \lesssim \sum_{r=1}^{R(\omega)} \frac{|p(\mathbf{s}^{(r)},\mathbf{c}^{(r)},\omega)|}{2^{\sum(s^{(r)}_j+c^{(r)}_j)}} \left[ s'^{(r)}_k \sqrt{(2s'^{(r)}_k-3)!!} \prod_{j\neq k} \sqrt{(2s'^{(r)}_j-1)!!} \right] \sigma^{S^{(r)}-1}.
\end{equation}

For the suppressed case ($s'_k = 0$), the derivative scales linearly with $\theta_k$, increasing the total order in $\sigma$:
\begin{equation}
\sqrt{\mathbb{E}[|\partial_{\theta_k} c_\omega|^2]} \lesssim \sum_{r=1}^{R(\omega)} \frac{|p(\mathbf{s}^{(r)},\mathbf{c}^{(r)},\omega)|}{2^{\sum(s^{(r)}_j+c^{(r)}_j)}} \left[ c'^{(r)}k \prod_{j\neq k} \sqrt{(2s'^{(r)}_j-1)!!} \right] \sigma^{S^{(r)}-s'_k+1}.
\end{equation}

For the standard case of no weight sharing ($s'_j=1, c'_j \in \{0,1\}$), the double factorials simplify ($(-1)!!=1$ and $1!!=1$), and the expression simplifies purely to the redundancy and scaling factors:
\begin{equation}
    \sqrt{\mathbb{E}[|\partial_{\theta_k} c_\omega|^2]} \lesssim \sum_{r=1}^{R(\omega)} \frac{|p(\mathbf{s}^{(r)},\mathbf{c}^{(r)},\omega)|}{2^{\sum(s^{(r)}_j+c^{(r)}_j)}} \sigma^{w-1}.
\end{equation}
This confirms that the second moment (RMS) follows the exact same $\sigma^{w-1}$ scaling law as the first moment (Mean Absolute), validating the tightness of the bounds used in the main theorem.

\subsection{Expected Upper Bound}\label{appx:exp_upper_bound_int}
\begin{proof}
We seek a rigorous upper bound on the expected magnitude of the loss gradient. We begin with the expression for the gradient of the loss at a specific frequency $\omega$, derived in Eq.~\eqref{eq:freq_loss_deriv}:
\begin{equation}
    \left| \partial_{\theta_k} L(\omega) \right| = \left| 2\text{Re}\left( c^*_{D_\omega} \partial_{\theta_k} c_{\omega_f} \right) \right|.
\end{equation}

First, we apply the inequality $|\text{Re}(z)| \leq |z|$ to bound the magnitude:
\begin{equation}
    \left| \partial_{\theta_k} L(\omega) \right| \leq 2 \left| c_{D_\omega}(\boldsymbol{\theta}) \right| \left| \partial_{\theta_k} c_{\omega_f}(\boldsymbol{\theta}) \right|.
\end{equation}
Next, we substitute the definition of the model error $c_{D_\omega} = c_{\omega_h} - c_{\omega_f}(\boldsymbol{\theta})$ and apply the triangle inequality $|A - B| \leq |A| + |B|$:
\begin{equation}
    \left| \partial_{\theta_k} L(\omega) \right| \leq 2 \left( |c_{\omega_h}| + |c_{\omega_f}(\boldsymbol{\theta})| \right) \left| \partial_{\theta_k} c_{\omega_f}(\boldsymbol{\theta}) \right|.
\end{equation}
We now take the expectation over the parameter distribution $\boldsymbol{\theta} \sim \mathcal{N}(0, \sigma^2 \mathbf{I})$. Using the linearity of the expectation operator, we split the bound into two distinct terms:
\begin{equation}
    \mathbb{E}\left[ \left| \partial_{\theta_k} L(\omega) \right| \right] \leq 2 |c_{\omega_h}| \mathbb{E}\left[ \left| \partial_{\theta_k} c_{\omega_f} \right| \right] + 2 \mathbb{E}\left[ \left| c_{\omega_f} \right| \left| \partial_{\theta_k} c_{\omega_f} \right| \right].
\end{equation}
 
Applying Cauchy-Schwarz:
\begin{equation}
    \mathbb{E}\left[ \left| \partial_{\theta_k} L(\omega) \right| \right] \leq 2 |c_{\omega_h}| \mathbb{E}\left[ \left| \partial_{\theta_k} c_{\omega_f} \right| \right] + 2 \sqrt{\mathbb{E}[|c_{\omega_f}|^2] \mathbb{E}[|\partial_{\theta_k} c_{\omega_f}|^2]}.
\end{equation}
We evaluate this bound by substituting the explicit moment sums derived over the $R(\omega)$ active variational paths. We consider the dominant regime where the differentiated parameter $\theta_k$ has sine-dependence ($s'_k \ge 1$), which provides the leading-order contribution.

\textit{First term:}
Substituting the explicit expression for the expected gradient magnitude eq..:
\begin{equation}
    2 |c_{\omega_h}| \mathbb{E}[|\partial_{\theta_k} c_{\omega_f}|] \lesssim 2 |c_{\omega_h}| \sum_{r=1}^{R(\omega)} \mathcal{C}^{(r)} \left[ \frac{s'^{(r)}_k 2^{\frac{s'^{(r)}_k-1}{2}}\Gamma(\frac{s'^{(r)}_k}{2})}{\sqrt{\pi}} \prod_{j\neq k} \frac{2^{\frac{s'^{(r)}_j}{2}}\Gamma(\frac{s'^{(r)}_j+1}{2})}{\sqrt{\pi}} \right] \sigma^{S^{(r)}-1},
\end{equation}
where $\mathcal{C}^{(r)} = |p(\mathbf{s}^{(r)},\mathbf{c}^{(r)},\omega)| \, 2^{-\sum(s^{(r)}_j+c^{(r)}_j)}$ encapsulates the combinatorial and trigonometric scaling factors for the $r$-th path.

\textit{Second term:}
This term involves the product of the RMS coefficient magnitude and the RMS gradient. The RMS coefficient magnitude scales as $\sigma^{S^{(r)}}$:
\begin{equation}
    \sqrt{\mathbb{E}[|c_{\omega_f}|^2]} \lesssim \sum_{r=1}^{R(\omega)} \mathcal{C}^{(r)} \left[ \prod_{j} \sqrt{(2s'^{(r)}_j-1)!!} \right] \sigma^{S^{(r)}}.
\end{equation}
The RMS gradient is given by the explicit second-moment sum:
\begin{equation}
    \sqrt{\mathbb{E}[|\partial_{\theta_k} c_{\omega_f}|^2]} \lesssim \sum_{r=1}^{R(\omega)} \mathcal{C}^{(r)} \left[ s'^{(r)}_k \sqrt{(2s'^{(r)}_k-3)!!} \prod_{j\neq k} \sqrt{(2s'^{(r)}_j-1)!!} \right] \sigma^{S^{(r)}-1}.
\end{equation}
Multiplying these two factors, the self-interaction term scales as $\sigma^{S^{(r)}} \times \sigma^{S^{(r)}-1} = \sigma^{2S^{(r)}-1}$.

Comparing the two components of the upper bound, we observe that the target-interaction term scales as $\sigma^{S-1}$ while the self-interaction term scales as $\sigma^{2S-1}$. In the small-angle initialization regime ($\sigma \ll 1$), the first term dominates. Thus, the expected gradient is bounded by the explicit sum:
\begin{equation}
    \mathbb{E}\left[ \left| \partial_{\theta_k} L(\omega) \right| \right] \lesssim 2 |c_{\omega_h}| \sum_{r=1}^{R(\omega)} \frac{|p(\mathbf{s}^{(r)},\mathbf{c}^{(r)},\omega)|}{2^{\sum(s^{(r)}_j+c^{(r)}_j)}} \left[ \frac{s'^{(r)}_k 2^{\frac{s'^{(r)}_k-1}{2}}\Gamma(\frac{s'^{(r)}_k}{2})}{\sqrt{\pi}} \prod_{j\neq k} \frac{2^{\frac{s'^{(r)}_j}{2}}\Gamma(\frac{s'^{(r)}_j+1}{2})}{\sqrt{\pi}} \right] \sigma^{S^{(r)}-1}.
\end{equation}
This result confirms that the gradient signal is a sum over $R(\omega)$ coherent paths. While the geometric prefactors (Gamma functions) grow with the sine-powers $s'_j$, they are dominated by the exponential suppression factor $\sigma^{S^{(r)}-1}$.

\textit{No weight sharing:}
In the case of with no weight sharing of variational parameters, we have $s'^{(r)}_j=1$ at most for all active parameters. The sine-degree $S^{(r)}$ becomes the variational depth $d_\omega$. The geometric prefactors simplify significantly:
\begin{equation}
    \frac{\Gamma(1/2)}{\sqrt{\pi}} = 1 \quad \text{and} \quad \frac{2^{1/2}\Gamma(1)}{\sqrt{\pi}} = \sqrt{\frac{2}{\pi}}.
\end{equation}
The bound reduces to the concise scaling law:
\begin{equation}
    \mathbb{E}\left[ \left| \partial_{\theta_k} L(\omega) \right| \right] \lesssim 2 |c_{\omega_h}| \sum_{r=1}^{R(\omega)} \frac{|p(\mathbf{s}^{(r)},\mathbf{c}^{(r)},\omega)|}{2^{\sum (s^{(r)}_j+c^{(r)}_j)}} \left( \frac{2}{\pi} \right)^{\frac{d_\omega-1}{2}} \sigma^{d_\omega-1}.
\end{equation}
In the standard case of Pauli encoding with fixed circuit depth $d$, the exponential suppression factor $\sigma^{d-1}$ applies uniformly across the spectrum. However, the result demonstrates that the expected gradient magnitude for a frequency $\omega$ is directly proportional to its redundancy $R(\omega)$ and the target coefficient magnitude $|c_{\omega_h}|$. This scaling establishes the mechanism of spectral bias through combinatorial redundancy.
\end{proof}

\subsection{Expected Non-Integer Upper Bound}\label{appx:exp_upper_bound_nonint}
\begin{proof}We extend the expected gradient analysis to the case of non-integer frequencies. Starting from the deterministic bound derived in Proof~\ref{proof:up_bound_nonint}:\begin{equation}|\partial_{\theta_k} L(\omega)| \leq \sum_{\omega'} |\text{sinc}(\pi(\omega-\omega'))| \left( \left| c^*{_{D_{\omega'}}} \right| \left| \partial_{\theta_k} c_{\omega_f} \right| + \left| c_{D_\omega} \right| \left| \partial_{\theta_k} c_{\omega'f} \right| \right).\end{equation}We take the expectation over the parameter distribution $\boldsymbol{\theta} \sim \mathcal{N}(0, \sigma^2 \mathbf{I})$. Using the linearity of expectation, we bring the operator inside the summation:\begin{equation}\mathbb{E}[|\partial{\theta_k} L(\omega)|] \leq \sum_{\omega'} |\text{sinc}(\pi(\omega-\omega'))| \left( \mathbb{E}\left[ |c_{D_{\omega'}}| |\partial_{\theta_k} c_{\omega_f}| \right] + \mathbb{E}\left[ |c_{D_\omega}| |\partial_{\theta_k} c_{\omega'f}| \right] \right).\end{equation}
We apply the same bounding strategy used for the integer case. For each interaction term $\mathbb{E}[|c_{D_\omega}||\partial c_{\omega_f}|]$, we use Cauchy-Schwarz and the triangle inequality to separate the target contribution from the model self-interaction. For small initialization variance $\sigma \ll 1$, the terms are dominated by the target coefficients $|h\omega|$ and $|h_{\omega'}|$. Neglecting the higher-order self-interaction terms ($\mathcal{O}(\sigma^{2d-1})$), the expectation simplifies to:\begin{equation}\mathbb{E}[|\partial_{\theta_k} L(\omega)|] \lesssim \sum_{\omega'} |\text{sinc}(\pi(\omega-\omega'))| \left( |h_{\omega'}| \mathbb{E}[|\partial_{\theta_k} c_{\omega_f}|] + |h_{\omega}| \mathbb{E}[|\partial_{\theta_k} c_{\omega'_f}|] \right).\end{equation}We now substitute the explicit moment sums derived in Sections~\ref{sec:exp_val} and~\ref{sec:second_moment}. For the dominant regime where the differentiated parameter has sine-dependence ($s'_k \ge 1$):

\textbf{1. Self Gradient Contribution ($\mathbb{E}[|\partial c_{\omega_f}|]$):}The gradient at frequency $\omega$ is a sum over its $R(\omega)$ active paths.\begin{equation}\mathbb{E}[|\partial_{\theta_k} c_{\omega_f}|] \lesssim \sum_{r=1}^{R(\omega)} \frac{|p(\mathbf{s}^{(r)},\mathbf{c}^{(r)},\omega)|}{2^{\sum(s^{(r)}_j+c^{(r)}_j)}} \left[ \frac{s'^{(r)}_k 2^{\frac{s'^{(r)}_k-1}{2}}\Gamma(\frac{s'^{(r)}k}{2})}{\sqrt{\pi}} \prod_{j\neq k} \frac{2^{\frac{s'^{(r)}_j}{2}}\Gamma(\frac{s'^{(r)}_j+1}{2})}{\sqrt{\pi}} \right] \sigma^{S^{(r)}-1}.\end{equation}\textbf{2. Leakage Gradient Contribution ($\mathbb{E}[|\partial c_{\omega'_f}|]$):}The gradient at the neighbor frequency $\omega'$ is a sum over its $R(\omega')$ active paths (indexed by $q$).\begin{equation}\mathbb{E}[|\partial_{\theta_k} c_{\omega'f}|] \lesssim \sum_{q=1}^{R(\omega')} \frac{|p(\mathbf{s}^{(q)},\mathbf{c}^{(q)},\omega')|}{2^{\sum(s^{(q)}_j+c^{(q)}_j)}} \left[ \frac{s'^{(q)}_k 2^{\frac{s'^{(q)}_k-1}{2}}\Gamma(\frac{s'^{(q)}k}{2})}{\sqrt{\pi}} \prod_{j\neq k} \frac{2^{\frac{s'^{(q)}_j}{2}}\Gamma(\frac{s'^{(q)}_j+1}{2})}{\sqrt{\pi}} \right] \sigma^{S^{(q)}-1}.\end{equation}
Substituting these explicit sums back into the main bound yields the final expression:

\begin{multline}\mathbb{E}[|\partial_{\theta_k} L(\omega)|] \lesssim \sum_{\omega'} |\text{sinc}(\pi(\omega-\omega'))| \Bigg( \|h_{\omega'}| \sum_{r=1}^{R(\omega)} \mathcal{G}^{(r)}(\omega) \sigma^{S^{(r)}-1}+ |h_{\omega}| \sum_{q=1}^{R(\omega')} \mathcal{G}^{(q)}(\omega') \sigma^{S^{(q)}-1} \Bigg),\end{multline}

where $\mathcal{G}^{(r)}(\omega)$ represents the explicit geometric and combinatorial prefactor for the $r$-th path of frequency $\omega$:
\begin{equation}\mathcal{G}^{(r)}(\omega) = \frac{|p(\mathbf{s}^{(r)},\mathbf{c}^{(r)},\omega)|}{2^{\sum(s^{(r)}_j+c^{(r)}_j)}} \left[ \frac{s'^{(r)}_k 2^{\frac{s'^{(r)}_k-1}{2}}\Gamma(\frac{s'^{(r)}k}{2})}{\sqrt{\pi}} \prod_{j\neq k} \frac{2^{\frac{s'^{(r)}_j}{2}}\Gamma(\frac{s'^{(r)}_j+1}{2})}{\sqrt{\pi}} \right].
\end{equation}

This result generalizes spectral bias to the non-integer regime. The expected gradient for a frequency $\omega$ is a weighted sum over the spectrum. Because the sinc function decays as $1/|\omega-\omega'|$, the gradient is primarily determined by three factors: the variational complexities $S^{(r)}$ and $S^{(q)}$ (which provide exponential suppression $\sigma^{S-1}$), the target amplitudes $|h|$, and crucially, the redundancies $R(\omega)$ and $R(\omega')$. Since the inner sums scale linearly with the number of contributing paths, frequencies with high redundancy or those in close proximity to high-redundancy neighbors receive significantly larger gradient updates. Thus, even with spectral leakage, high-complexity frequencies remain exponentially suppressed unless they benefit from the redundancy of a nearby low-complexity, high-amplitude target frequency.
\end{proof}
\section{Extended redundancy analysis results}\label{appx:redundancy_analysis_extended}

\begin{figure}[t!]
    \centering
    \begin{subfigure}[b]{0.49\linewidth}
        \includegraphics[width=\linewidth]{figs/redundancy_analysis_constant.pdf}
        \caption{Constant Pauli encoding}
        \label{fig:redundancy_a}
    \end{subfigure}
    \hfill
    \begin{subfigure}[b]{0.49\linewidth}
        \includegraphics[width=\linewidth]{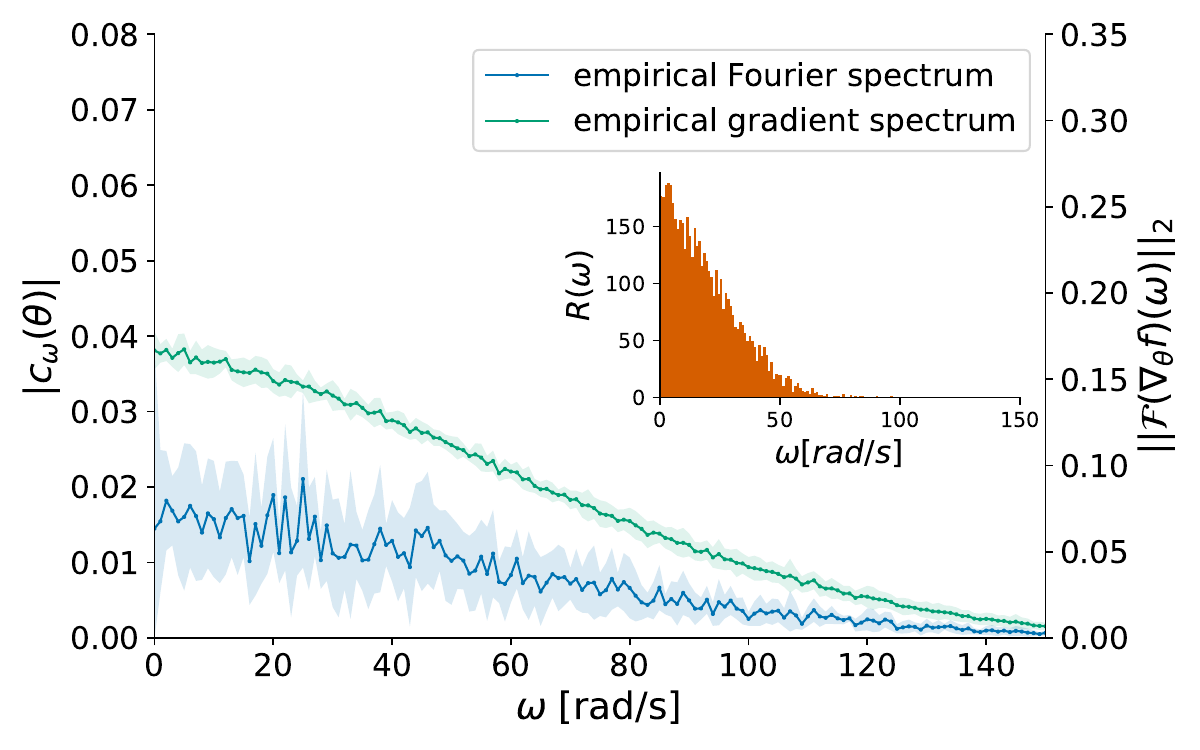}
        \caption{Linear Pauli encoding}
        \label{fig:redundancy_b}
    \end{subfigure}
    \begin{subfigure}[b]{0.49\linewidth}
        \includegraphics[width=\linewidth]{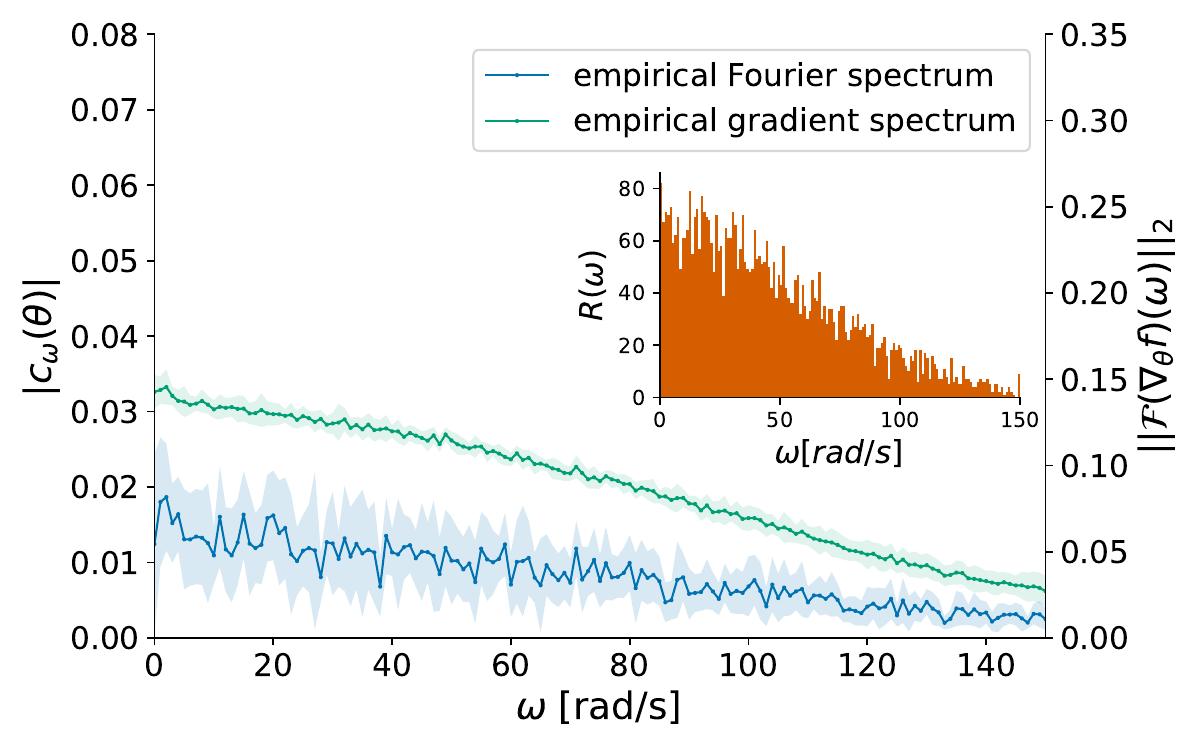}
        \caption{Exponential Pauli encoding}
        \label{fig:redundancy_c}
    \end{subfigure}
    \hfill
    \begin{subfigure}[b]{0.49\linewidth}
        \includegraphics[width=\linewidth]{figs/redundancy_analysis_exp.pdf}
        \caption{Exponential Pauli encoding}
        \label{fig:redundancy_c}
    \end{subfigure}

    \caption{The Fourier spectra of the four Pauli encoding schemes with empirical data taken as the sampled mean over ten models. Depicted are the sampled theoretically accessible frequencies (light red), the mean Fourier coefficient (blue), and the total gradient of trainable parameters at each Fourier coefficient (green).}
    \label{fig:redundancy_extended}
\end{figure}

An extended set of results depicting the spectra of PQCs with a variety of encodings in Figure~\ref{fig:redundancy_extended}. The encodings are constant Pauli ($\beta_i = 1$), linear Pauli ($\beta_i = i$), binary Pauli ($\beta_i = 2^i$) and exponential Pauli ($\beta_i = 3^i$).

\section{Extended Spectral bias results}\label{appx:spectral_bias_extended}

\begin{figure}[t!]
    \centering
    \begin{subfigure}[b]{0.49\linewidth}
        \includegraphics[width=\linewidth]{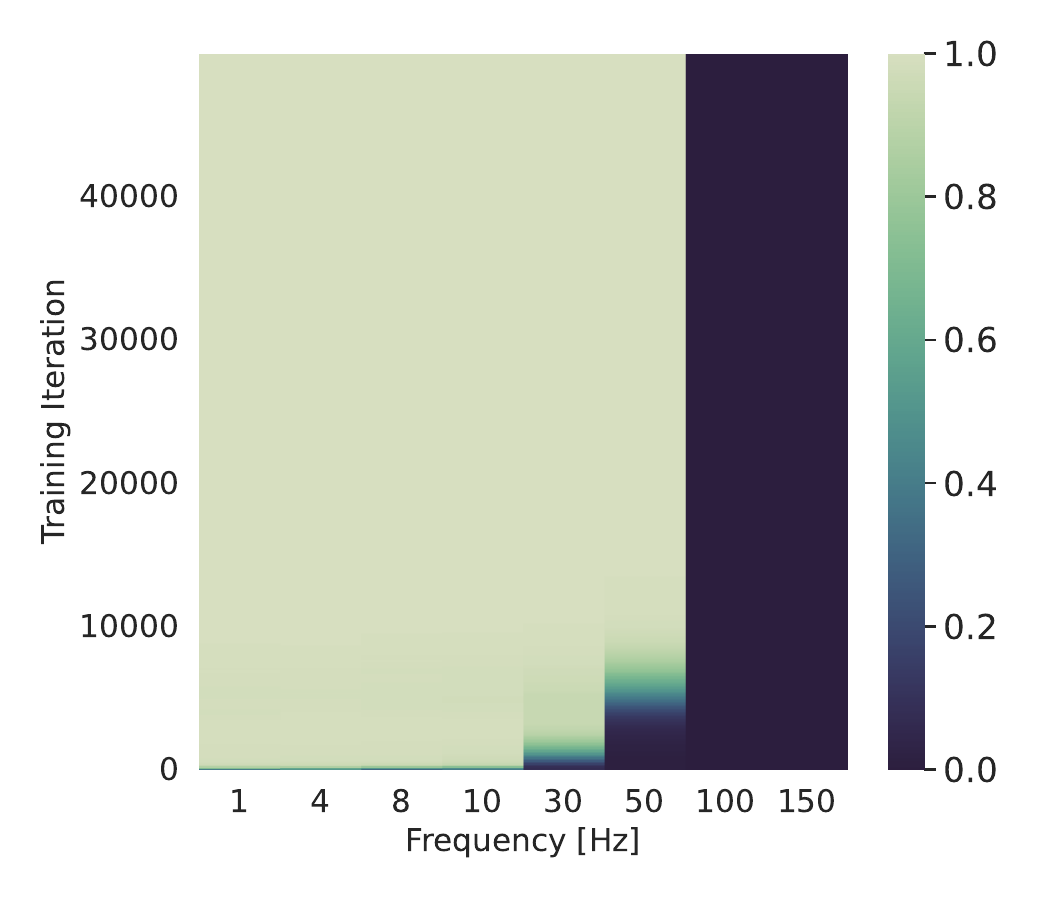}
        \caption{Constant Pauli encoding}
    \label{fig:spectral_bias_a}
    \end{subfigure}
    \hfill
    \begin{subfigure}[b]{0.49\linewidth}
        \includegraphics[width=\linewidth]{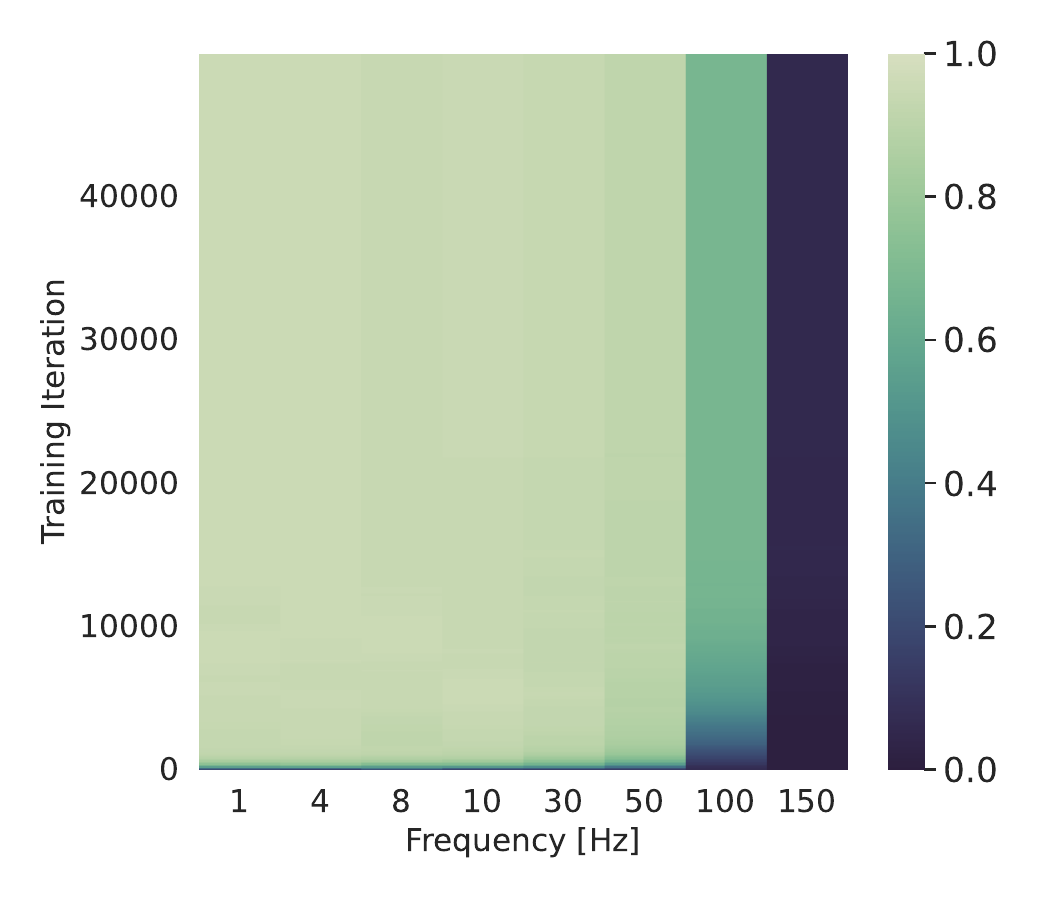}
        \caption{Linear Pauli encoding}
    \label{fig:spectral_bias_b}
    \end{subfigure}
    \begin{subfigure}[b]{0.49\linewidth}
        \includegraphics[width=\linewidth]{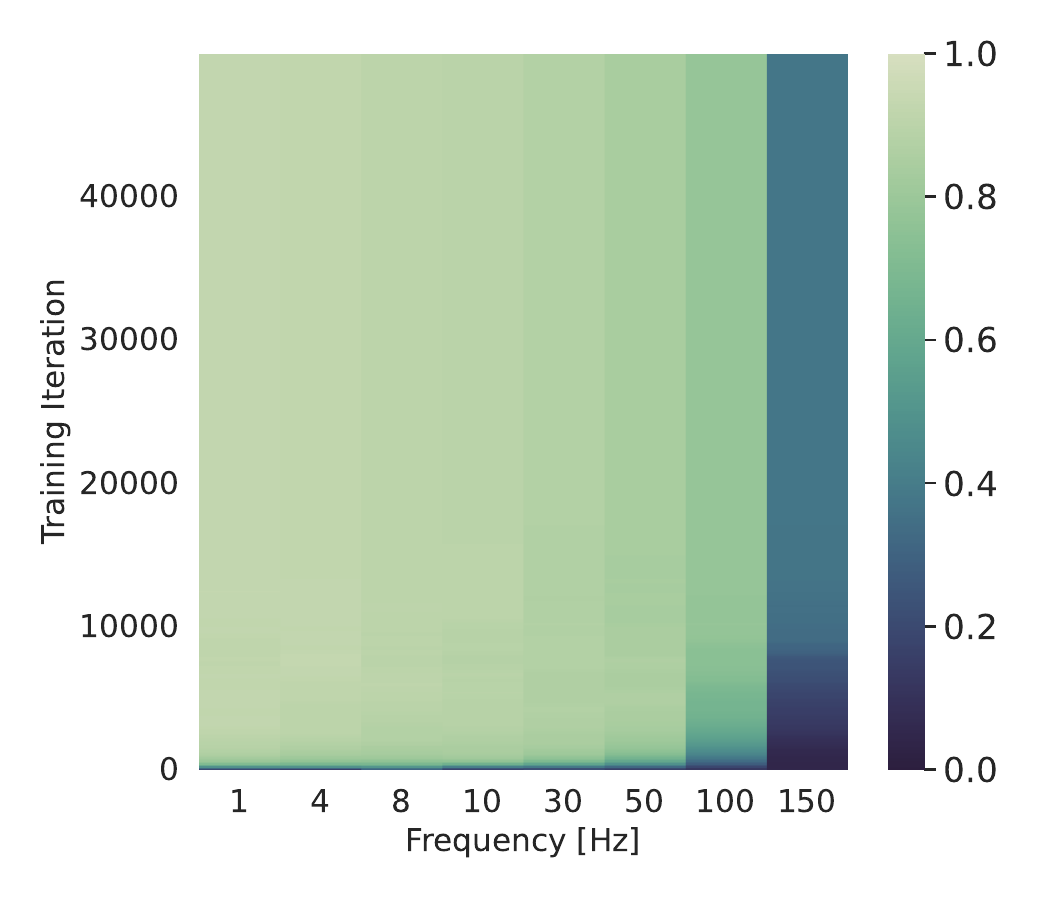}
        \caption{Binary Pauli encoding}
    \label{fig:spectral_bias_c}
    \end{subfigure}
\hfill
            \begin{subfigure}[b]{0.49\linewidth}
        \includegraphics[width=\linewidth]{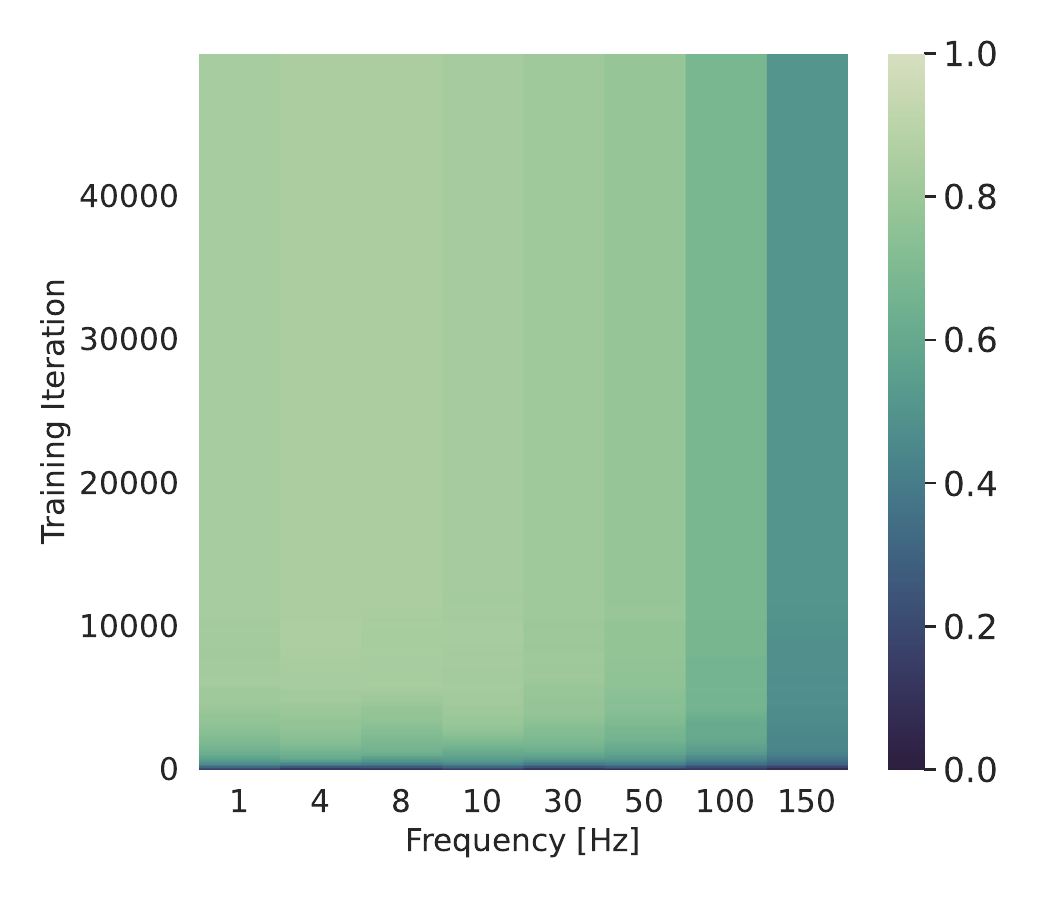}
        \caption{Exponential Pauli encoding}
    \label{fig:spectral_bias_b}
    \end{subfigure}

    \caption{The rate at which frequencies (x-axis) are learnt during the course of training (y-axis), the colorbar measures the PQC spectrum normalised by the target amplitude at a given frequency ($|\tilde{f}_\omega|/A_i$). Each subplot depicts the training dynamics for a different encoding scheme (a) constant Pauli encoding, (b) linear Pauli encoding and (c) exponential Pauli encoding.}
    \label{fig:spectral_bias_extended}
\end{figure}

An extended set of results depicting the spectra of PQCs with a variety of encodings in Figure~\ref{fig:spectral_bias_extended}. The encodings are constant Pauli ($\beta_i = 1$), linear Pauli ($\beta_i = i$), binary Pauli ($\beta_i = 2^i$) and exponential Pauli ($\beta_i = 3^i$).

\section{Robustness of redundant Fourier coefficients under parameter perturbation}\label{appx:robustness_proof}
Here we detail the proof of theorem \ref{thm:robustness}.
\paragraph{Notation and setup.}
For a fixed frequency \(\omega\). Decompose the (learned) Fourier coefficient with trained parameters \(\theta^*\) as a sum of \(R(\omega)\) contributing terms.
\[
c_\omega(\theta)=\sum_{i=1}^{R(\omega)} a_i(\theta^*),
\]
where each \(a_i(\theta^*)\in\mathbb{C}\). Let \(\delta\theta\) be a random additive perturbation, and define
\[
\Delta a_i := a_i(\theta^*+\delta\theta)-a_i(\theta^*),\qquad
\Delta C := \sum_{i=1}^{R(\omega)} \Delta a_i.
\]
We consider the normalized RMS deviation (root-mean-square fractional deviation)
\[
\mathcal{R} \;:=\; \frac{\sqrt{\mathbb{E}[\,|\Delta C|^2\,]}}{|c(\theta^*)|},
\]
and we assume the coefficient of interest is non-zero \(c_\omega(\theta^*)\).

\paragraph{Assumptions.}
\begin{enumerate}
  \item[(A1)] (\emph{Partial coherence}.) There exist constants \(\bar a>0\) and \(\kappa\in(0,1]\) such that
  \[
  |c_{\omega}(\theta^*)| \ge \kappa\, R(\omega)\,\bar a,
    \qquad \bar a := \frac{1}{R(\omega)}\sum_{i=1}^{R(\omega)} |a_i(\theta^*)|.
  \]

  \item[(A2)] (\emph{Second-moment and pairwise covariance control}.) For the perturbation distribution,
  \[
  \mathbb{E}\big[|\Delta a_i|^2\big] = \sigma_a^2 \quad\text{for all }i,
  \]
  and for \(i\ne j\),
  \[
  \big|\mathbb{E}[\Delta a_i\,\overline{\Delta a_j}]\big|\le \rho\,\sigma_a^2
  \]
  for some \(\rho\in[0,1]\).
  \item[(A3)] (\emph{Perturbation scale dependence}.) The variance \(\sigma^2_a=\sigma^2_a(\|\delta\theta\|)\) depends only on the perturbation magnitude; for small perturbations one typically has \(\sigma_a\propto\|\delta\theta\|\) by linearisation.
\end{enumerate}

\begin{proof}
Compute the second moment:
\[
\mathbb{E}[|\Delta C|^2] \;=\; \mathbb{E}\Big[\Big|\sum_{i=1}^R \Delta a_i\Big|^2\Big]
\;=\; \sum_{i=1}^R \mathbb{E}[|\Delta a_i|^2] + \sum_{i\ne j}\mathbb{E}[\Delta a_i\overline{\Delta a_j}].
\]
Applying the uniform bounds in (A2) the first sum equals \(R\sigma_a^2\) and each off-diagonal term has magnitude at most \(\rho\sigma_a^2\). Taking absolute values and bounding,
\[
\mathbb{E}[|\Delta C|^2] \le R(\omega)\sigma_a^2 + R(\omega)(R(\omega)-1)\rho\sigma_a^2 = R(\omega)\sigma_a^2\big(1+(R(\omega)-1)\rho\big).
\]
Hence
\[
\sqrt{\mathbb{E}[|\Delta C|^2]} \le \sigma_a\sqrt{R(\omega)\big(1+(R(\omega)-1)\rho\big)}.
\]
Using (A1) to lower bound the denominator,
\[
\mathcal{R} \le \frac{\sigma_a\sqrt{R(\omega)(1+(R(\omega)-1)\rho)}}{\kappa R(\omega) \bar a}
= \frac{\sigma_a}{\kappa\bar a}\sqrt{\frac{1+(R(\omega)-1)\rho}{R(\omega)}},
\]
which is \eqref{eq:general-bound}.
\end{proof}

If \(\rho\to 0\) (approximate independence of perturbation effects across summands) the bound reduces to the \(1/\sqrt{R(\omega)}\) scaling up to multiplicative constants. If \(\rho\) is close to \(1\) (highly correlated perturbation effects), the redundancy offers no benefit.

\end{document}